\documentclass[journal]{IEEEtran}
\usepackage{cite}
\usepackage{amsmath,amssymb,amsfonts}
\usepackage{algorithmic}
\usepackage{graphicx}
\usepackage{caption,setspace}
\usepackage{etoolbox}
\usepackage{xfrac}
\usepackage{multirow}

\ifCLASSINFOpdf
\else
\fi
\usepackage{amsmath}
\usepackage{xfrac}
\DeclareMathOperator*{\argmin}{argmin}

\usepackage[cmintegrals]{newtxmath}
\begin{document}
\title{Improved User Fairness in Decode-Forward Relaying Non-orthogonal Multiple Access Schemes with Imperfect SIC}

\author{Ferdi~Kara,~\IEEEmembership{ Member,~IEEE,}
        Hakan~Kaya~
\thanks{ This work is supported by the Scientific and Technological Research Council of Turkey (TUBITAK) under the 2211-E program}
\thanks{The authors are with Wireless Communication Technologies Laboratory (WCTLab), Department
of Electrical and Electronics Engineering, Zonguldak Bülent Ecevit University, Zonguldak, 67100, Turkey  (e-mail: \{f.kara,hakan.kaya\}@beun.edu.tr)}}
\maketitle

\begin{abstract}
Non-orthogonal multiple access (NOMA) is one of the key technologies to serve in ultra-dense networks with massive connections which is crucial for Internet of Things (IoT) applications. Besides, NOMA provides better spectral efficiency compared to orthogonal multiple access (OMA) schemes. However, in NOMA, successive interference canceler (SIC) should be implemented for interference mitigation and mostly in the literature, perfect SIC is assumed for NOMA involved systems. Unfortunately, this is not the case for practical scenarios and this imperfect SIC effect limits the performance of NOMA involved systems. In addition, it causes unfairness between users. In this paper, we introduce reversed decode-forward relaying NOMA (R-DFNOMA) to improve user fairness compared to conventional DFNOMA (C-DFNOMA) which is widely analyzed in literature. In the analysis, we define imperfect SIC effect dependant to channel fading and with this imperfect SIC,  we derive exact expressions for ergodic capacity (EC) and outage probability (OP). Moreover, we evaluate bit error performance of proposed R-DFNOMA and derive bit error probability (BEP) in closed-form which has not been also studied well in literature. Then, we define user fairness index in terms of all key performance indicators (KPIs) (i.e., EC, OP and BEP). Based on extensive simulations, all derived expressions are validated, and it is proved that proposed R-DFNOMA provides better user fairness than C-DFNOMA in terms of all KPIs. Finally, we discuss the effect of power allocations at the both source and relay on the performance metrics and user fairness.
\end{abstract}
%
\section{Introduction}
%
%
%
%
\IEEEPARstart{I}{n} recent years, exponential increase in connected devices (i.e., smart-phones, tablets, watches etc.) \cite{VNI} to the internet and with the introduce of the Internet of Things (IoT), future radio networks (FRN) are keen to serve massive users in dense networks which is called Massive Machine Type Communication (mMTC) -one of the three major concepts of 5G and beyond- \cite{Andrews2014}. Non-orthogonal Multiple Access (NOMA) is seen as a strong candidate for mMTC in FRN due to its high spectral efficiency and ability to support massive connections \cite{Shirvanimoghaddam2017}. In NOMA, users are assigned into same resource block to increase spectral efficiency and the most attracted scheme is power domain (PD)-NOMA where users share the same resource block with different power allocation coefficients \cite{Dai2015}. The interference mitigation in PD-NOMA is held by successive interference canceler (SIC) \cite{Saito2013}. Due to its potential for 5G and beyond, NOMA\footnote{NOMA is used for PD-NOMA after this point.} has attracted tremendous attention from researchers where NOMA is widely investigated mostly in terms of achievable rate/ergodic capacity (EC) and outage probability (OP) \cite{6868214}. Besides, since only superposition coding at the transmitter and SICs at the receivers are required, the integration of NOMA with other physical layer techniques such as cooperative communication, mm-wave communication, multi-input-multi-output (MIMO) systems, visible light communication, etc. has also taken a remarkable attention \cite{Ding2017}.
\subsection{Related Works and Motivation}
One of the most attracted topics is the interplay between NOMA and cooperative communication which is held in three major concepts: 1) cooperative-NOMA \cite{7117391,Kara2019} where near users also act as relays for far user, 2) NOMA-based cooperative systems \cite{Kim2015a,Kara2020a} where NOMA is implemented to increase spectral efficiency of device-to-device communication and 3) relay-assisted/aided-NOMA where relays in the network help NOMA users to enhance coverage. This paper focuses on relay-assisted-NOMA networks. The relay-assisted-NOMA networks have also been analyzed widely. In these works, an amplify-forward (AF) or a decode-forward (DF) relay helps source/BS to transmit symbols to the NOMA users \cite{8287018}. Hybrid DF/AF relaying strategies have been investigated to improve outage performance of relay-assisted-NOMA \cite{7556297}. Outage and sum-rate performance of relay-assisted-NOMA networks have also been analyzed whether a direct link between the source and the users exists \cite{Liu2018} or not \cite{8466778,7870605,7876764}. Then, relay-assisted-NOMA networks have been analyzed in terms of achievable rate and outage performance under different conditions such as buffer aided-relaying \cite{7747506,7803598}, partial channel state information (CSI) at transmitter \cite{8103803} and imperfect CSI at receiver \cite{7752764} when a single relay is located between the source and the users. In addition, relay selection schemes have been investigated when multiple relays are available \cite{7482785,8038031,8329423,8031883}. Relay selection schemes are based on guaranteeing QoS of users and maximizing outage performance of users. Moreover, two-way relaying strategies where relay operates as a coordinated multi-point (CoMP), have been investigated in terms of achievable rate and outage performance \cite{7417453,7819537,8275033,8316931}.

However, in aforementioned either conventional or relay-assisted NOMA networks, mostly perfect SIC is assumed. This is not a reasonable assumption when considered fading channels. To the best of the authors' knowledge, very limited studies investigate NOMA involved system with imperfect SIC. However, in those works, the imperfect SIC effect is assumed to be independent from the channel fading \cite{7819537,8275033, Im2019}. Thus, this strict assumption should be relaxed. Besides, all the studies with imperfect SIC \cite{7819537,8275033, Im2019} have been devoted to two-way relaying NOMA systems. To the best of the authors' knowledge, relay assisted NOMA networks have not been analyzed with imperfect SIC effects. Moreover, once the imperfect SIC is taken into consideration, it is shown that in downlink NOMA schemes, users encounter a performance degradation in bit/symbol error rate (BER/SER) compared to orthogonal multiple access (OMA) though its performance gains in terms of EC and OP \cite{Kara2018,Assaf2019}. Indeed, this performance degradation may be more severe for one of the users. Hence, the user fairness should be also considered in system design. Although this is raised in conventional downlink NOMA networks \cite{Timotheou2015} and some studies are devoted to improve user fairness in conventional NOMA networks in terms of EC and OP \cite{Liu2015a,Liu2016,Liu2016e}, to the best of the authors' knowledge, user fairness in terms of BER/SER in conventional NOMA has been taken into consideration. Moreover, this user unfairness becomes worse in relay-assisted-NOMA systems due to the effects of two phases (e.g., from source-to-relay and from relay-to-users).  Besides all this, BER performance of relay-assisted-NOMA networks has been only analyzed in \cite{Kara2020b} though they have been widely-analyzed in terms of EC and OP. User fairness also has not been considered for relay-assisted-NOMA networks in terms of any key performance indicators (KPIs) (e.g., EC, OP, BER). To this end, we analyze relay-assisted-NOMA network with imperfect SIC for all performance metrics. The user fairness has been also raised for relay-assisted-NOMA networks.

\subsection{Contributions}
The main contributions of this paper are as follow:
\begin{itemize}
\item{We introduce reversed DF relaying NOMA (R-DFNOMA) to improve user fairness in conventional DFNOMA (C-DFNOMA).}
\item{For a more realistic/practical scenario, we re-define imperfect SIC effect as dependant to channel fading coefficient. The capacity and outage performances of proposed R-DFNOMA are investigated with this imperfect SIC effect. The exact EC expressions are derived and closed-form upper bounds are provided for EC. Besides, exact OP expressions are derived in closed-forms. All derived expressions match perfectly with simulations.}
\item{Contrary to the most of the literature, we also analyze the error performance of R-DFNOMA rather than only EC and/or OP performances.
Exact bit error probability (BEP) expressions are provided in closed-forms and validated via computer simulations.}
\item{We evaluate the performances of proposed model in terms of all KPIs (i.e., EC, OP and BEP) and compared with the benchmark. In this content, to the best of the authors' knowledge, this is also the first study which provides an overall performance evaluation for any NOMA involved systems. All literature researches have biased on investigations for only one or two performance metrics (e.g., EC and/or OP).}
\item{We define users fairness in terms of all KPIs (i.e., EC, OP and BEP). Based on extensive simulations, it is proved that proposed R-DFNOMA provides better user fairness compared to C-DFNOMA. Finally, we reveal the effect of power allocation on user fairness and discuss optimum power allocation}\end{itemize}
\subsection{Organization}
The remainder of this paper is as follows. In Section II, the proposed R-DFNOMA and the benchmark C-DFNOMA schemes are introduced. The detection algorithms at the users and the signal-to-interference plus noise ratio (SINR) definitions are also provided in this section. Then, the performance analysis for three KPIs (i.e., EC, OP, BER) are derived in Section III and the user fairness indexes for all KPIs are provided. In Section IV, all derived expressions are validated via Monte Carlo simulations. In addition, performance comparisons are also revealed in this section. Finally, results are discussed and the paper is concluded in Section V.
\subsection{Notation}
The list of symbols, notations and abbreviations through this paper is given in Table 1.
\section{System and Channel Model}
\subsection{Proposed: Reversed DF Relaying in NOMA}
\begin{figure}[t]
  \centering
  \includegraphics[width=8cm]{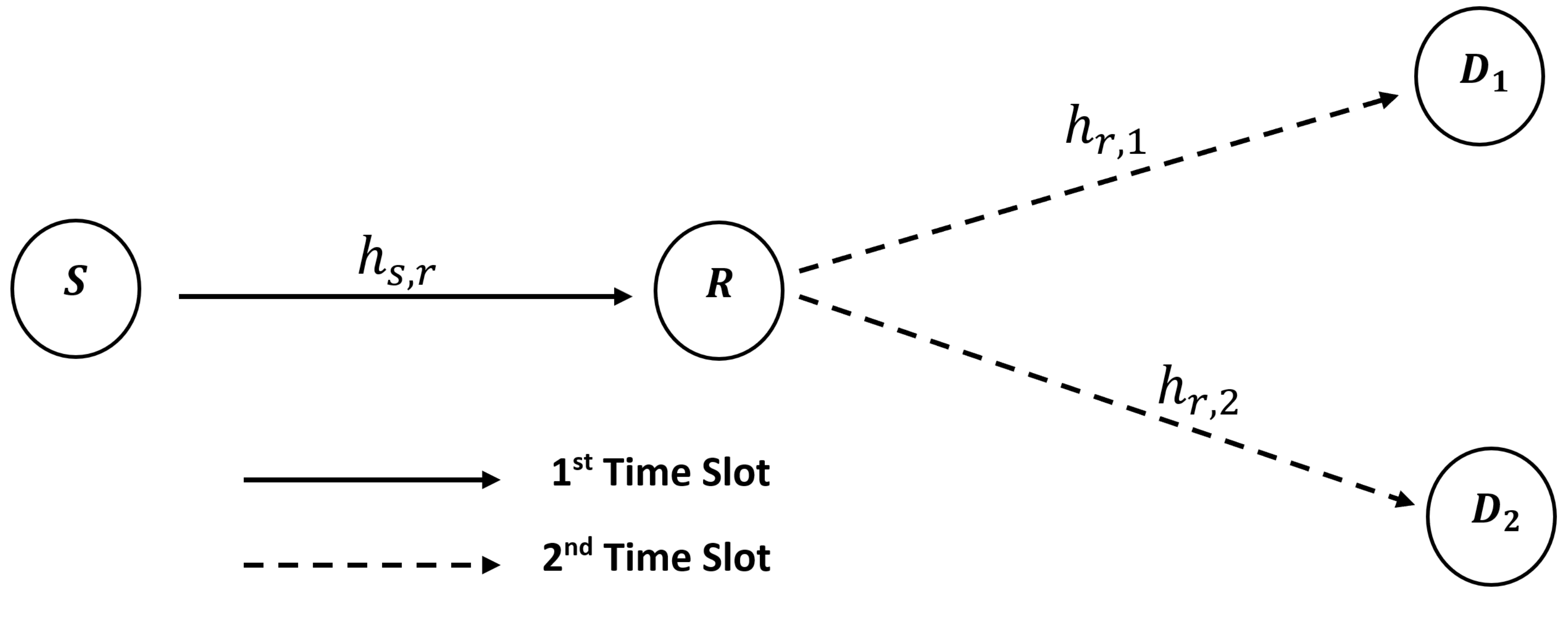}
  \caption{The illustration of R-DFNOMA}
  \label{fig1}
\end{figure}
As shown in Fig. 1, a source ($S$) communicates with two destinations (i.e., $D_1$ and $D_2$) with the help of a relay ($R$). The relay applies decode-forward (DF) strategy in a half-duplex mode, thus the total communication occupies two time slots. We assume that direct links from source to destinations are not available due to the high path-loss effects and/or obstacles. According to their average channel qualities between relay and destinations (i.e., $R-D_1$ and $R-D_2$, users are defined as near and far users. We assume that $D_1$ has a better channel than $D_2$ to the relay node ($R$). In this case, $D_1$ and $D_2$ are denoted as near and far user, respectively and the system design is handled. In the first phase of communication, source ($S$) implements superposition coding for the base-band modulated symbols of destinations (i.e., $x_1$ and $x_2$) and transmits it to the relay. The received signal by the relay is given as
\begin{table}[!t]
\centering
\caption{List of Symbols, Notations and Abbreviations}
\begin{tabular}{|l|l|}
\hline
$P_i$ & Transmit power at node $i=s,r$\\ \hline
$\alpha_j$ & Power allocation at the source for user $j$ \\ \hline
$x_j$ &Modulated base-band (IQ) symbol of user $j$ \\ \hline
\multirow{2}{*}{$h_{i,k}$} & Flat fading channel coefficient \\
& between nodes $i$ and $k$ \\ \hline
$n_i$ &Additive white Gaussian noise (AWGN) \\ \hline
$\mu$&Propagation constant \\ \hline
$\tau$ &Path loss exponent \\ \hline
$d_{i,k}$ &Euclidean distance between $i$ and $k$ \\ \hline
$\sim$& Follows/distributed \\ \hline
\multirow{2}{*}{$CN(m,\sigma^2)$}& Complex Normal distribution with $m$ mean \\
&and $\frac{\sigma^2}{2}$ variance in each component \\ \hline
$\left|.\right|$& Absolute value \\ \hline
$\rho_i$ & Transmit signal-to-noise ratio (SNR) \\ \hline
\multirow{2}{*}{$SINR_j^{i,k}$} & Signal-to-interference plus noise ratio (SINR)  \\
&for user $j$ between nodes $i$ and $k$ \\ \hline
\multirow{2}{*}{$\gamma_{i,k}$} & Absolute square for channel fading between \\
&nodes $i$ and $k$ ($\left|h_{i,k}\right|^2$)\\ \hline
\multirow{2}{*}{$\hat{x}_j$} & Detected/estimated base-band (IQ) symbol \\
& of user $j$ at relay \\ \hline
$\Xi_i$ &Imperfect SIC effect coefficient at the node $i$ \\ \hline
$\beta_j$ & Power allocation at the relay for user $j$ \\ \hline
\multirow{2}{*}{$\tilde{x}_j$} & Detected/estimated base-band (IQ) symbol of \\
&user $j$ at destination \\ \hline
$R_j$ &Achievable (Shannon) rate of user $j$ \\ \hline
$C_j$ & Ergodic capacity (EC) of user $j$ \\ \hline
\multirow{2}{*}{$f_Z(z)$} & Probability density function (PDF) \\
& of random variable  $z$ \\ \hline
\multirow{2}{*}{$F_Z(z)$} & Cumulative distribution function (CDF) \\
&of random variable  $z$ \\ \hline
$\hat{R}_j$ &Target rate of user $j$ (QoS requirement) \\ \hline
$P_j(out)$ & Outage probability (OP) of user $j$ \\ \hline
$e2e$&End-to-end \\ \hline
\multirow{2}{*}{$P_j^{(i,k)}(e)$} & Bit error probability (BEP) of user $j$  \\
& between nodes $i$ and $k$ \\ \hline
$P_j(e|\gamma_{i,k})$ & Conditional BEP on  $\gamma_{i,k}$ \\ \hline
\multirow{2}{*}{$Q(.)$} & Marcum-Q function \\
& $Q(z)=\int\limits_z^\infty\sfrac{1}{\sqrt{2\pi}}exp(-\sfrac{z^2}{2})dz$ \\ \hline
\multirow{2}{*}{$A_j^{(i,k)}(e)$} & Coefficient $A$ of user $j$  between nodes $i$ and $k$ \\
&in BEP analysis \\ \hline
\multirow{2}{*}{$PF_l(out)$} & Proportional fairness index  \\
& $l=c,o,e$ $c\to$EC, $o\to$OP and $e\to$BEP \\ \hline
\end{tabular}
\end{table}

\begin{equation}
y_r=\sqrt{P_s}h_{s,r}\left(\sqrt{\alpha_1}x_1+\sqrt{\alpha_2}x_2\right)+n_r
\end{equation}
where $P_s$ is the transmit power of source. $h_{s,r}$ and $n_r$ denote the complex flat fading channel coefficient between $S-R$ and the additive white Gaussian noise (AWGN) at the relay. They follow $CN\left(0,\sigma_{s,r}^2\right)$ and $CN\left(0,N_0\right)$, respectively. $\sigma_{s,r}^2$ includes the large-scale fading effects and $\sigma_{s,r}^2=\mu d_{s,r}^{-\tau}$ is defined where $\mu$ and $\tau$ are the propagation constant and path-loss exponent, respectively. $d_{s,r}$ is the Euclidean distance between the nodes. In (1), $\alpha_1$ and $\alpha_2$ are the power allocation coefficient for the symbol of $D_1$ and $D_2$, respectively. In order to improve user fairness, in R-DFNOMA, we propose to allocate $\alpha_1>\alpha_2$ in the first phase where $\alpha_1+\alpha_2=1$. Unlike previous works, we propose to reverse power allocation coefficient in the first phase (e.g., $\alpha_1>\alpha_2$ ) and conventional power allocation is proposed in the second phase (e.g., $\beta_2>\beta_1$ -will be defined below-) whereas in conventional DFNOMA schemes, they have performed same way in both phases -as defined in benchmark in the next subsection-. Thus, the proposed system model is called as reversed-DFNOMA (R-DFNOMA). This reversed power allocation brings also reversed detecting order in the first phase. Since more power is allocated to $D_1$ symbols, relay node ($R$) firstly detects $x_1$ symbols by pretending $x_2$ symbols as noise based on the received signal in the first phase. The maximum-likelihood (ML) detection of $x_1$ symbols at the relay is given
\begin{equation}
  \hat{x}_1=\argmin_{k}{\left|y_{s,r}-\sqrt{P_s}h_{s,r}\sqrt{\alpha_1}x_{1,k}\right|^2}
\end{equation}
where $x_{1,k}$ denotes the $k$ th point in the $M_1$-ary constellation. The received signal-to-interference plus noise ratio (SINR) for the $x_1$ symbols at the relay is given by
\begin{equation}
  SINR_1^{(s,r)}=\frac{\rho_s\alpha_1\gamma_{s,r}}{\rho_s\alpha_2\gamma_{s,r}+1}
\end{equation}
where $\rho_s=\sfrac{P_s}{N_0}$ and $\gamma_{s,r}=\left|h_{s,r}\right|^2$ are defined. On the other hand, a successive interference canceler (SIC) should be implemented at the relay to detect less-powered $x_2$ symbols. The ML detection of $x_2$ symbols at the relay is given as
\begin{equation}
  \hat{x}_2=\argmin_{k}{\left|y_{s,r}^{'}-\sqrt{P_s}h_{s,r}\sqrt{\alpha_2}x_{2,k}\right|^2}
\end{equation}
where
\begin{equation}
  y_{s,r}^{'}= y_{s,r}-\sqrt{P_s}h_{s,r}\sqrt{\alpha_1}\hat{x}_1
\end{equation}
and $x_{2,k}$ denotes the $k$ th point in the $M_2$-ary constellation. One can easily see that, the remaining signal after SIC highly depends on the detection of $x_1$ symbols and unlike previous works, it is not reasonable to assume perfect SIC (e.g., no interference from $x_1$ symbols). In addition, the interference after SIC is a function of $\gamma_{s,r}$, $P_s$ and $\alpha_1$, thus the interference cannot be assumed an independent random variable unlike given in \cite{7819537,8275033, Im2019}. To this end, the SINR for $x_2$ symbols at the relay is given as
\begin{equation}
  SINR_2^{(s,r)}=\frac{\rho_s\alpha_2\gamma_{s,r}}{\Xi_{r}\rho_s\alpha_1\gamma_{s,r}+1}
\end{equation}
where $\Xi_{r}$ defines the imperfect SIC effect coefficient (e.g., $\Xi_{r}=0$ for perfect SIC and $\Xi_{r}=1$ for no SIC at all).

In the second phase of communication, relay node (R) again implements superposition coding for detected $\hat{x}_1$ and $\hat{x}_2$ symbols and broadcasts this total symbol to the destinations. The received signal by both destinations is given as
\begin{equation}
y_i=\sqrt{P_r}h_{r,i}\left(\sqrt{\beta_1}\hat{x}_1+\sqrt{\beta_2}\hat{x}_2\right)+n_i \quad i=1,2
\end{equation}
where $P_r$ is the transmit power of relay \footnote{The relay can harvest its energy from received RF signal to transmit signals. However, in this paper this constraint has not been regarded and energy harvesting (EH) models such as linear and non-linear seen as future researches.}. $h_{r,i}$ and $n_i$ denote the complex flat fading channel coefficient between $R-D_i$ and the additive white Gaussian noise (AWGN) at the $D_i$. $h_{r,i}$ $\sim$ $CN\left(0,\sigma_{r,i}^2\right)$ and  $n_i$ $\sim$ $CN\left(0,N_0\right)$, respectively. We assume $d_{r,2}\geq d_{r,1}$, hence $D_1$ has better channel condition and more power allocated to $D_2$ -user with weaker channel condition-, (e.g., $\beta_2>\beta_1$). Based on received signals, users implement whether ML or SIC plus ML in order to detect their own symbols. Since more power is allocated for the symbols of $D_2$, $D_2$ implements only ML by pretending $D_1$'s symbols as noise and it is given,
\begin{equation}
\tilde{x}_2=\argmin_{k}{\left|y_{r,2}-\sqrt{P_2}h_{r,2}\sqrt{\beta_2}x_{2,k}\right|^2}.
\end{equation}
The received SINR at the $D_2$ is given as
\begin{equation}
SINR_2^{(r,2)}=\frac{\rho_r\beta_2\gamma_{r,2}}{\rho_r\beta_1\gamma_{r,2}+1}.
\end{equation}
where $\rho_r=\sfrac{P_r}{N_0}$ is defined.

On the other hand, $D_1$ implements SIC in order to detect its own symbols. Thus, it firstly detects $x_2$ symbols and subtract regenerated forms from received signal. The detection process at the $D_1$ is given as
\begin{equation}
  \tilde{x}_1=\argmin_{k}{\left|y_{r,1}^{'}-\sqrt{P_r}h_{r,1}\sqrt{\beta_1}x_{1,k}\right|^2}
\end{equation}
where
\begin{equation}
  y_{r,1}^{'}= y_{r,1}-\sqrt{P_r}h_{r,1}\sqrt{\beta_2}\tilde{x}_2.
\end{equation}

The received SINR after SIC at the $D_1$ is given as
\begin{equation}
  SINR_1^{(r,1)}=\frac{\rho_r\beta_1\gamma_{r,1}}{\Xi_{1}\rho_r\beta_2\gamma_{r,1}+1}
\end{equation}
where $\Xi_{1}$ defines the imperfect SIC effect coefficient at the $D_1$ likewise in relay.

\subsection{Benchmark: Conventional DF Relaying in NOMA}
In conventional DF relay-aided NOMA (C-DFNOMA) schemes, detecting order at both relay and user are the same. The power allocation in the first phase is arranged as $\alpha_2^{*}>\alpha_1^{*}$. Hence, the relay node (R) firstly detects $x_2$ symbols and implements SIC to detect $x_1$ symbols. To this end, given detection algorithms and SINR definitions eq. (2)-(6) should be re-defined. The detection of $x_2$  symbols at the relay is given
\begin{equation}
  \hat{x}_2=\argmin_{k}{\left|y_{s,r}-\sqrt{P_s}h_{s,r}\sqrt{\alpha_2^{*}}x_{2,k}\right|^2}
\end{equation}
and of $x_1$ symbols
\begin{equation}
  \hat{x}_1=\argmin_{k}{\left|y_{s,r}^{+}-\sqrt{P_s}h_{s,r}\sqrt{\alpha_1^{*}}x_{1,k}\right|^2}
\end{equation}
where
\begin{equation}
  y_{s,r}^{+}= y_{s,r}-\sqrt{P_s}h_{s,r}\sqrt{\alpha_2^{*}}\hat{x}_2.
\end{equation}
The SINRs in the first phase of communication are given as
\begin{equation}
  SINR_2^{(s,r)}=\frac{\rho_s\alpha_2^{*}\gamma_{s,r}}{\rho_s\alpha_1^{*}\gamma_{s,r}+1}
\end{equation}
and
\begin{equation}
  SINR_1^{(s,r)}=\frac{\rho_s\alpha_1^{*}\gamma_{s,r}}{\Xi_{r}\rho_s\alpha_2^{*}\gamma_{s,r}+1}.
\end{equation}

The signal detections and the SINRs in the second phase of C-DFNOMA are the same in R-DFNOMA.

\section{Performance Analysis}
In this section, we analyze the proposed R-DFNOMA in terms of three KPIs (i.e., EC, OP and BEP) in order to evaluate its performance. Then, we define user fairness index for all three KPIs.
\subsection{Ergodic Capacity (EC)}
Since the proposed model includes a relaying strategy, its achievable rate is limited by the weakest link. Hence, considering both $S-R$ and $R-D_1$ links, the achievable (Shannon) rate of $D_1$ is given as
\begin{equation}
R_1=\frac{1}{2}\min{\left\{\log_2{\left(1+SINR_1^{(s,r)}\right)},\log_2{\left(1+SINR_1^{(r,1)}\right)}\right\}}.
\end{equation}
where $\frac{1}{2}$ exists since the total communication covers two time slots. The ergodic capacity (EC) of $D_1$ is obtained by averaging $R_1$ over instantaneous SINRs in (3) and (12). It is given as
\begin{equation}
\begin{split}
&C_1=\frac{1}{2} \\
&\iint\limits_0^\infty \log_2{\left(1+\min{\left\{SINR_1^{(s,r)},SINR_1^{(r,1)}\right\}}\right)} f_{\gamma_{s,r}}f_{\gamma_{r,1}}d\gamma_{s,r}d\gamma_{r,1}
\end{split}
\end{equation}
where $f_{\gamma_{s,r}}$ and  $f_{\gamma_{r,1}}$ are probability density functions (PDFs) of $\gamma_{s,r}$ and $\gamma_{r,1}$, respectively. Let define $Z=\min{\left\{X,Y\right\}}$, the cumulative density function (CDF) of $Z$ is given by $F_Z(z)=1-\left(1-F_X(z)\right)\left(1-F_Y(z)\right)$ where $F_X(.)$ and $F_Y(.)$ are CDFs of $X$ and $Y$, respectively \cite{Devore2002}. Recalling, $\int\limits_0^\infty \log_2{(1+x)}f_X(x)dx=\frac{1}{ln2}\int\limits_0^\infty \frac{1-F_X(x)}{1+x}dx$ \cite{Gradshteyn1994}, with some algebraic manipulations, we derive EC of $D_1$ as
\begin{equation}
\begin{split}
C_1=\frac{1}{2ln2}\int\limits_0^\infty \frac{exp\left(-\frac{z}{\left(\alpha_1-\alpha_2z\right)\rho_s\sigma_{s,r}^2}-\frac{z}{\left(\beta_1-\beta_2\Xi_1z\right)\rho_r\sigma_{r,1}^2}\right)}{1+z}dz
\end{split}
\end{equation}

To the best of the authors' knowledge, (20) cannot be solved in closed-form analytically. Nevertheless, it can be easily computed by numerical tools. In addition, we can obtain it in the closed-form for high SNR regime. To this end, we assume that $\rho_s,\rho_r\to\infty$. In this case, $Z=\min{\left\{SINR_1^{(s,r)},SINR_1^{(r,1)}\right\}}$ in (19) turns out to be $\lim_{\rho_s,\rho_r\to\infty}{Z}=min{\left\{\frac{\alpha_1}{\alpha_2},\frac{\beta_1}{\Xi_1\beta_2}\right\}}$. With some algebraic simplifications, the upper bound for EC of $D_1$ is given by
\begin{equation}
C_1\approx\frac{1}{2}\log_2{\eta_1}
\end{equation}
where $\eta_1=min{\left\{\frac{\alpha_1}{\alpha_2},\frac{\beta_1}{\Xi_1\beta_2}\right\}}$.

Likewise in capacity analysis of $D_1$, the achievable rate of $D_2$ is given by
\begin{equation}
R_2=\frac{1}{2}\min{\left\{\log_2{\left(1+SINR_2^{(s,r)}\right)},\log_2{\left(1+SINR_2^{(r,2)}\right)}\right\}}.
\end{equation}
and taking the similar steps between (19)-(20), EC of $D_2$ is derived as
\begin{equation}
\begin{split}
C_2=\frac{1}{2ln2}\int\limits_0^\infty \frac{exp\left(-\frac{z}{\left(\alpha_2-\alpha_1\Xi_rz\right)\rho_s\sigma_{s,r}^2}-\frac{z}{\left(\beta_2-\beta_1z\right)\rho_r\sigma_{r,2}^2}\right)}{1+z}dz.
\end{split}
\end{equation}
Likewise (20), (21) can be easily computed by numerical tools. Again in order to obtain upper bound for EC of $D_2$, if we assume $\rho_s,\rho_r\to\infty$, the EC is obtained as
\begin{equation}
C_2\approx\frac{1}{2}\log_2{\eta_2}
\end{equation}
where $\eta_2=min{\left\{\frac{\alpha_2}{\Xi_r\alpha_1},\frac{\beta_2}{\beta_1}\right\}}$.
\subsection{Outage Probability (OP)}
The outage event for any user is defined as
\begin{equation}
P_i(out)=P\left(R_i<\acute{R}_i\right) \quad i=1,2
\end{equation}
where $\acute{R}_i$ is the target rate of $D_i$. By substituting (18) and (22) into (25), OPs of users are derived as
\begin{equation}
\begin{split}
&P_i(out)=\\
&P\left(\frac{1}{2}\min{\left\{\log_2{\left(1+SINR_i^{(s,r)}\right)},\log_2{\left(1+SINR_i^{(r,i)}\right)}\right\}}<\acute{R}_i\right) \\
&\quad \quad\quad\quad\quad \quad \quad \quad\quad \quad \quad \quad\quad \quad \quad \quad\quad \quad \quad \quad  i=1,2.
\end{split}
\end{equation}
With some algebraic manipulations, OPs of users are derived as
\begin{equation}
P_i(out)=F_{Z_i}(\phi_i) \quad i=1,2
\end{equation}
where $\phi_i=2^{2\acute{R}_i}-1$ and $F_{Z_i}(.)$ CDF of $Z_i=\min{\left\{SINR_i^{(s,r)},SINR_i^{(r,i)}\right\}} \quad i=1,2$ are defined. Recalling CDF for minimum of two exponential random variables in (20) and (23), OP of users are derived in the closed-forms as
\begin{equation}
\begin{split}
&P_1(out)=\\
&1-exp\left(-\frac{\phi_1}{\left(\alpha_1-\alpha_2\phi_1\right)\rho_s\sigma_{s,r}^2}-\frac{\phi_1}{\left(\beta_1-\beta_2\Xi_1\phi_1\right)\rho_r\sigma_{r,1}^2}\right)
\end{split}
\end{equation}
and
\begin{equation}
\begin{split}
&P_2(out)=\\
&1-exp\left(-\frac{\phi_2}{\left(\alpha_2-\alpha_1\Xi_r\phi_2\right)\rho_s\sigma_{s,r}^2}-\frac{\phi_2}{\left(\beta_2-\beta_1\phi_2\right)\rho_r\sigma_{r,2}^2}\right).
\end{split}
\end{equation}
\subsection{Bit Error Probability (BEP)}
Since a cooperative communication is included in R-DFNOMA, the number of total erroneous bits from source to destination (i.e., end-to-end (e2e)) of users are given as
\begin{equation}
\begin{split}
  &N_i=\\
  &N_i\left(x_i\rightarrow\hat{x}_i\right)+N_i\left(\hat{x}_i\rightarrow\tilde{x}_i\right)-N_i\left(x_i\rightarrow\hat{x}_i\right)\cap N_i\left(\hat{x}_i\rightarrow\tilde{x}_i\right),\\
   &\quad \quad\quad\quad\quad \quad \quad \quad\quad \quad \quad \quad\quad \quad \quad \quad\quad \quad \quad \quad i=1,2
\end{split}
\end{equation}
where $N_i\left(x_i\rightarrow\hat{x}_i\right)$ and $N_i\left(\hat{x}_i\rightarrow\tilde{x}_i\right)$ denote the number of erroneous detected bits of $D_i$ in the first and second phases, respectively. If erroneous detections have been performed in both phase, this means that correct detection has been achieved from source to destinations (e2e). Thus, the set of intersection of erroneous detections (3rd term) is subtracted in (30). Considering all combinations, the BEPs of $D_i$ are given as in (31) (see top of the next page).
\begin{figure*}
\begin{equation}
\begin{split}
P_i(e)=\frac{1}{M_i\log_2{M_i}}
\left[\sum_{k=1}^{M_i}\sum_{\forall l\neq k}\sum_{\forall p\neq l}N_i\left(x_{i,k}\rightarrow\hat{x}_{i,l}\right)+N_i\left(\hat{x}_{i,l}\rightarrow\tilde{x}_{i,p}\right)-N_i\left(x_{i,k}\rightarrow\hat{x}_{i,l}\right)\cap N_i\left(\hat{x}_{i,l}\rightarrow\tilde{x}_{i,p}\right)\right] &\\
\quad i=1,2&
\end{split}
\end{equation}
\hrulefill
\end{figure*}

Recalling that $N_i\left(x_i\rightarrow\hat{x}_i\right)$ and $N_i\left(\hat{x}_i\rightarrow\tilde{x}_i\right)$ events are statistically independent, thus with the law of total probability, BEPs of users are given as
\begin{equation}
\begin{split}
&P_i^{(e2e)}(e)=\\
&P_i^{(s,r)}(e)\left(1-P_i^{(r,i)}(e)\right)+\left(1-P_i^{(s,r)}(e)\right)P_i^{(r,i)}(e) \quad i=1,2
\end{split}
\end{equation}
where $P_i^{(s,r)}(e)=\frac{1}{M_i\log_2{M_i}}\sum_{k=1}^{M_i}\sum_{\forall l\neq k}N_i\left(x_{i,k}\rightarrow\hat{x}_{i,l}\right)$ and $P_i^{(r,i)}(e)=\frac{1}{M_i\log_2{M_i}}\sum_{l=1}^{M_i}\sum_{\forall p\neq l}N_i\left(\hat{x}_{i,l}\rightarrow\tilde{x}_{i,p}\right)$ denote the BEPs in the first and second phases, respectively. Thus, the BEPs in each phases should be firstly derived. Each phase of communication can be considered separately. In the first phase of communication, it turns out to be a conventional downlink NOMA system and the BEPs of $x_1$ symbols will be the same with BEP of \textit{far user} in downlink NOMA. Since the superposition is applied, the BEP of \textit{far user} in NOMA is highly depended on the chosen constellation pairs (i.e., $M_1$ and $M_2$)\cite{Kara2018,Assaf2019}. Nevertheless, the conditional BEP on channel conditions is given in the form,
\begin{equation}
P_{1}^{(s,r)}(e|_{\gamma_{s,r}})=\sum_{q=1}^{L_1^{(s,r)}}\varsigma_{1,q}^{(s,r)}Q\left(\sqrt{2\nu_{1,q}^{(s,r)}\rho_s\gamma_{s,r}}\right),
\end{equation}
where $L_{1}^{(s,r)}$, $\varsigma_{1,q}^{(s,r)}$ and $\nu_{1,q}^{(s,r)}$ coefficients change according to chosen modulation constellation pairs for $x_1$ and $x_2$ symbols \cite[Table 1]{Kara2019a}. For instance, in case $M_1=M_2=4$ is used for both symbols (i.e., $x_1$ and $x_2$), $L_{1}^{(s,r)}=2$, $\varsigma_{1,q}^{(s,r)}=0.5 \ \forall q$ and $\nu_{1,q}^{(s,r)}=\frac{1}{2} \left(\sqrt{\alpha_1}\mp\sqrt{\alpha_2}\right)^2$ (for proof see \cite[Appendix A]{Kara2019c}). Then, recalling $\gamma_{s,r}$ is exponentially distributed, with the aid of \cite{Alouini1999} the average BEP (ABEP) of $x_1$ symbols in the first phase is obtained as,
\begin{equation}
P_{1}^{(s,r)}(e)=\sum_{q=1}^{L_1^{(s,r)}}\frac{\varsigma_{1,q}^{(s,r)}}{2}\left(1-\sqrt{\frac{\nu_{1,q}^{(s,r)}\rho_s\sigma_{s,r}^2}{1+\nu_{1,q}^{(s,r)}\rho_s\sigma_{s,r}^2}}\right).
\end{equation}
On the other hand, $x_2$ symbols in the first phase can be considered as \textit{near user symbols} in conventional downlink NOMA, Thus, the conditional BEP should be derived considering correct and erroneous SIC cases. After summing these BEPs of two cases, the conditional BEP of $x_2$ symbols in the first phase is given in the form just as (33)
\begin{equation}
P_{2}^{(s,r)}(e|_{\gamma_{s,r}})=\sum_{q=1}^{L_2^{(s,r)}}\varsigma_{2,q}^{(s,r)}Q\left(\sqrt{2\nu_{2,q}^{(s,r)}\rho_s\gamma_{s,r}}\right),
\end{equation}
where $L_{2}^{(s,r)}=5$, $\varsigma_{2,q}^{(s,r)}=\frac{1}{2}[2, 1, -1, -1, 1]$ and $\nu_{2,q}^{(s,r)}=[\frac{\alpha_1}{2}, \frac{\left(\sqrt{\alpha_2}-\sqrt{\alpha_1}\right)^2}{2}, \frac{\left(\sqrt{\alpha_2}+\sqrt{\alpha_1}\right)^2}{2}, \frac{\left(2\sqrt{\alpha_2}-\sqrt{\alpha_1}\right)^2}{2}, \frac{\left(2\sqrt{\alpha_2}+\sqrt{\alpha_1}\right)^2}{2}]$ are given for $M_1=M_2=4$ \cite[Appendix A and B]{Kara2019c}. By averaging over instantaneous $\gamma_{s,r}$, the ABEP of $x_2$ symbols in the first phase is derived as
\begin{equation}
P_{2}^{(s,r)}(e)=\sum_{q=1}^{L_2^{(s,r)}}\frac{\varsigma_{2,q}^{(s,r)}}{2}\left(1-\sqrt{\frac{\nu_{2,q}^{(s,r)}\rho_s\sigma_{s,r}^2}{1+\nu_{2,q}^{(s,r)}\rho_s\sigma_{s,r}^2}}\right).
\end{equation}

In the second phase of communication, more power is allocated to $\hat{x}_2$ symbols. Thus, $D_2$ implements a ML detection without SIC so the BEP of $x_2$ symbols in the second phase can be easily derived by using (33) as
\begin{equation}
P_{2}^{(r,2)}(e|_{\gamma_{r,2}})=\sum_{t=1}^{L_2^{(r,2)}}\varsigma_{2,q}^{(r,2)}Q\left(\sqrt{2\nu_{2,q}^{(r,2)}\rho_r\gamma_{r,2}}\right),
\end{equation}
where $L_2^{(r,2)}\triangleq L_1^{(s,r)}$, $\varsigma_{2,q}^{(r,2)}\triangleq\varsigma_{1,q}^{(s,r)}$ and $\nu_{2,q}^{(r,2)}\triangleq\nu_{1,q}^{(s,r)}$. By using (34), (36), the ABEP is given as
\begin{equation}
P_{2}^{(r,2)}(e)=\sum_{q=1}^{L_2^{(r,2)}}\frac{\varsigma_{2,q}^{(r,2)}}{2}\left(1-\sqrt{\frac{\nu_{2,q}^{(r,2)}\rho_r\sigma_{r,2}^2}{1+\nu_{2,q}^{(r,2)}\rho_r\sigma_{r,2}^2}}\right).
\end{equation}
Likewise, the BEP of $x_1$ symbols in the second phase can be easily obtained by repeating steps (35), (35). The conditional BEP and the ABEP are given as
\begin{equation}
P_{1}^{(r,1)}(e|_{\gamma_{r,1}})=\sum_{q=1}^{L_1^{(r,1)}}\varsigma_{1,q}^{(r,1)}Q\left(\sqrt{2\nu_{1,q}^{(r,1)}\rho_r\gamma_{r,1}}\right),
\end{equation}
and
\begin{equation}
P_{1}^{(r,1)}(e)=\sum_{q=1}^{L_1^{(r,1)}}\frac{\varsigma_{1,q}^{(r,1)}}{2}\left(1-\sqrt{\frac{\nu_{1,q}^{(r,1)}\rho_r\sigma_{r,1}^2}{1+\nu_{1,q}^{(r,1)}\rho_r\sigma_{r,1}^2}}\right).
\end{equation}
where $L_1^{(r,1)}\triangleq L_2^{(s,r)}$, $\varsigma_{1,q}^{(r,1)}\triangleq\varsigma_{2,q}^{(s,r)}$ and $\nu_{1,q}^{(r,1)}\triangleq\nu_{2,q}^{(s,r)}$.

Lastly, substituting (34), (36), (38) and (40) into (32), the ABEPs of users are derived as in (41) and (42) (see top of the next page).
\begin{figure*}
\begin{equation}
\begin{split}
 P_{1}^{(e2e)}(e)=&\sum_{q=1}^{L_1^{(s,r)}}\frac{\varsigma_{1,q}^{(s,r)}}{2}\left(1-\sqrt{\frac{\nu_{1,q}^{(s,r)}\rho_s\sigma_{s,r}^2}{1+\nu_{1,q}^{(s,r)}\rho_s\sigma_{s,r}^2}}\right)\left[1-\sum_{q=1}^{L_1^{(r,1)}}\frac{\varsigma_{1,q}^{(r,1)}}{2}\left(1-\sqrt{\frac{\nu_{1,q}^{(r,1)}\rho_r\sigma_{r,1}^2}{1+\nu_{1,q}^{(r,1)}\rho_r\sigma_{r,1}^2}}\right)\right]\\
 &+\left[1-\sum_{q=1}^{L_1^{(s,r)}}\frac{\varsigma_{1,q}^{(s,r)}}{2}\left(1-\sqrt{\frac{\nu_{1,q}^{(s,r)}\rho_s\sigma_{s,r}^2}{1+\nu_{1,q}^{(s,r)}\rho_s\sigma_{s,r}^2}}\right)\right]\sum_{q=1}^{L_1^{(r,1)}}\frac{\varsigma_{1,q}^{(r,1)}}{2}\left(1-\sqrt{\frac{\nu_{1,q}^{(r,1)}\rho_r\sigma_{r,1}^2}{1+\nu_{1,q}^{(r,1)}\rho_r\sigma_{r,1}^2}}\right)
\end{split}
\end{equation}
\hrulefill
\end{figure*}
\begin{figure*}
\begin{equation}
\begin{split}
 P_{2}^{(e2e)}(e)=&\sum_{q=1}^{L_2^{(s,r)}}\frac{\varsigma_{2,q}^{(s,r)}}{2}\left(1-\sqrt{\frac{\nu_{2,q}^{(s,r)}\rho_s\sigma_{s,r}^2}{1+\nu_{2,q}^{(s,r)}\rho_s\sigma_{s,r}^2}}\right)\left[1-\sum_{q=1}^{L_2^{(r,2)}}\frac{\varsigma_{2,q}^{(r,2)}}{2}\left(1-\sqrt{\frac{\nu_{2,q}^{(r,2)}\rho_r\sigma_{r,2}^2}{1+\nu_{2,q}^{(r,2)}\rho_r\sigma_{r,2}^2}}\right)\right]\\
 &+\left[1-\sum_{q=1}^{L_2^{(s,r)}}\frac{\varsigma_{2,q}^{(s,r)}}{2}\left(1-\sqrt{\frac{\nu_{2,q}^{(s,r)}\rho_s\sigma_{s,r}^2}{1+\nu_{2,q}^{(s,r)}\rho_s\sigma_{s,r}^2}}\right)\right]\sum_{q=1}^{L_2^{(r,2)}}\frac{\varsigma_{2,q}^{(r,2)}}{2}\left(1-\sqrt{\frac{\nu_{2,q}^{(r,2)}\rho_r\sigma_{r,2}^2}{1+\nu_{2,q}^{(r,2)}\rho_r\sigma_{r,2}^2}}\right)
\end{split}
\end{equation}
\hrulefill
\end{figure*}
\subsection{User Fairness}
In this subsection, we define fairness between users' performances. In NOMA schemes, since the total power is allocated between users, the users have different performances. Due to the inter-user-interference and the SIC operation, one of the users may have better performance than the other. This performance gap can be higher in some performance metrics (e.g. EC and BER).

The performance gap between users should not be increased. We use proportional fairness (PF) index to compare users' performances for all KPIs. For instance, let we firstly consider EC. In this case, if the fairness has not been considered, one of the users may achieve much more EC than the other. To alleviate this unfair situation, PF index for EC should be defined and it is given as
\begin{equation}
PF_c=\frac{C_1}{C_2}
\end{equation}
which can be easily obtained by substituting (20) and (23) into (43). One can easily see that optimum value for $PF_c$ can be considered as 1 which means that both user have exactly the same EC. Nevertheless, this may not be achieved when the users have different QoS requirements. Thus, fairness index should be obtained for other KPIs and all three should be evaluated together. To this end, fairness indexes for outage and error performances are given as
 \begin{equation}
PF_o=\frac{P_1(out)}{P_2(out)}
\end{equation}
and
 \begin{equation}
PF_e=\frac{P_1(e)}{P_2(e)}
\end{equation}
which can be computed by substituting (28), (29) into (44) and (41), (42) into (45), respectively. It is again clear that the optimal values for $PF_o$ and $PF_e$ are also 1. However likewise in $PF_c$, it may not be always achieved due to the priority in QoS requirements of users. It is noteworthy that in the PF index for all KPIs, $\kappa$ and $\sfrac{1}{\kappa}$  have the same meaning. For instance, if the PF index for any performance metric has $2$ and/or $0.5$, this means that one of the users has two times better performance than the other.

\section{Performance Evaluation}
In this section, we provide validation of the provided analysis in the previous sections. In addition, we present user fairness comparisons between proposed R-DFNOMA and C-DFNOMA\footnote{In C-DFNOMA, power allocation in the first phase is complement of the power allocation of R-DFNOMA (i.e., $\alpha_1^{*}=1-\alpha_1)$}. In all simulations, we assume that $\mu=10$ and $\tau=2$. The transmit power of source and relay are assumed to be equal (i.e., $P_s=P_r$). In validations of R-DFNOMA, unless otherwise stated, curves denote theoretical analysis\footnote{In numerical integration for exact EC, the infinity in the upper bounds of the integrals is changed with $10^3$ not to cause numerical calculation errors.} and simulations are demonstrated by markers. Moreover, in all simulations, the imperfect SIC effect coefficients at the both nodes are assumed to be equal (i.e., $\Xi_r=\Xi_1$).

\subsection{The Effect of Imperfect SIC}
In this subsection, the distances between the nodes are assumed to be $d_{s,r}=5$, $d_{r,1}=1$ and $d_{r,2}=3$. It can be seen from following figures that all derived expressions match perfectly with simulations.

\begin{figure}[!ht]
  \centering
  \includegraphics[width=8cm]{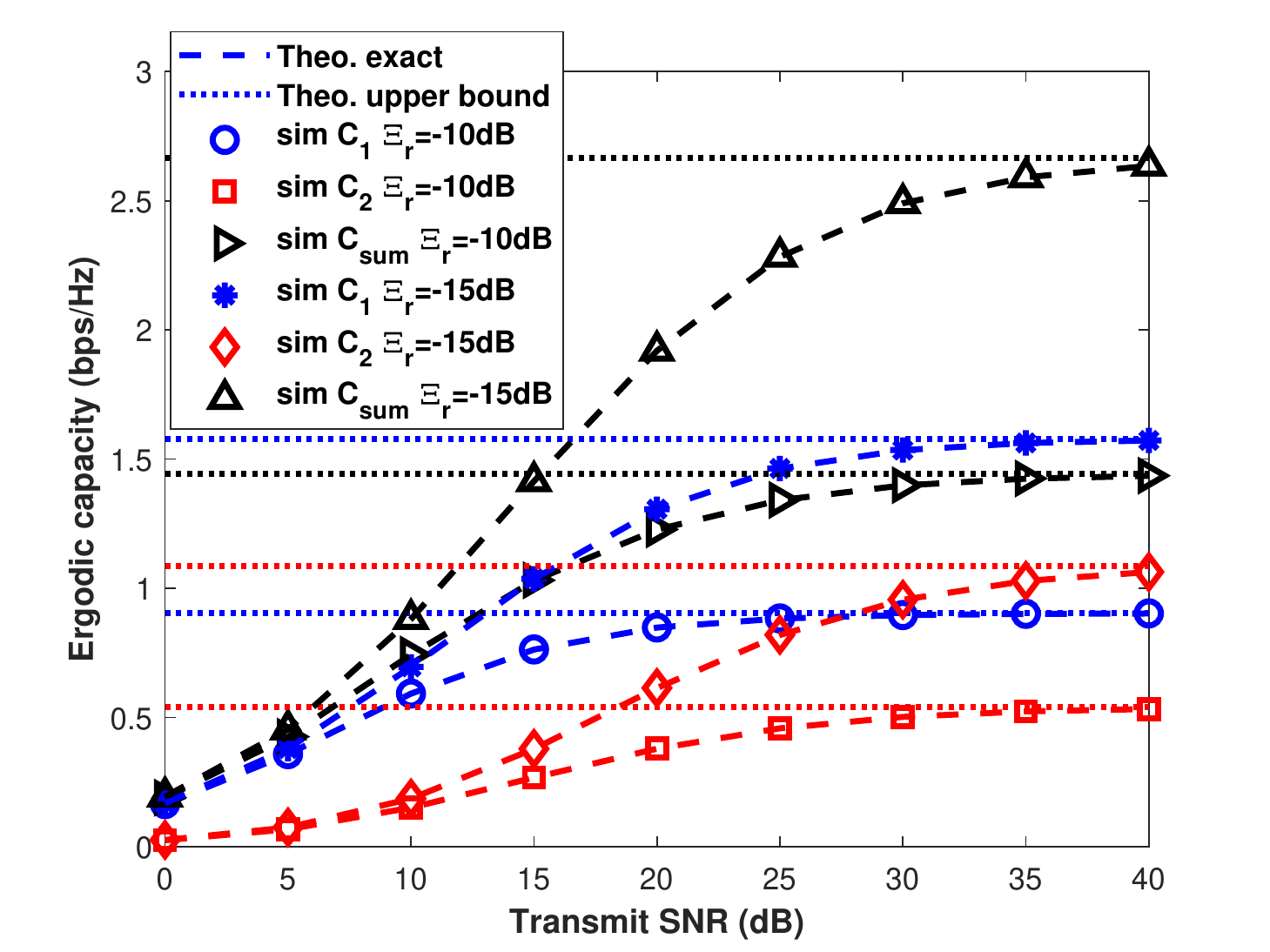}
  \caption{EC of R-DFNOMA vs $\rho_s$ when $\alpha_1=0.9$, $\beta_1=0.2$,  $d_{s,r}=5$, $d_{r,1}=1$ and $d_{r,2}=3$}
  \label{fig2}
\end{figure}
\begin{figure}[!ht]
  \centering
  \includegraphics[width=8cm]{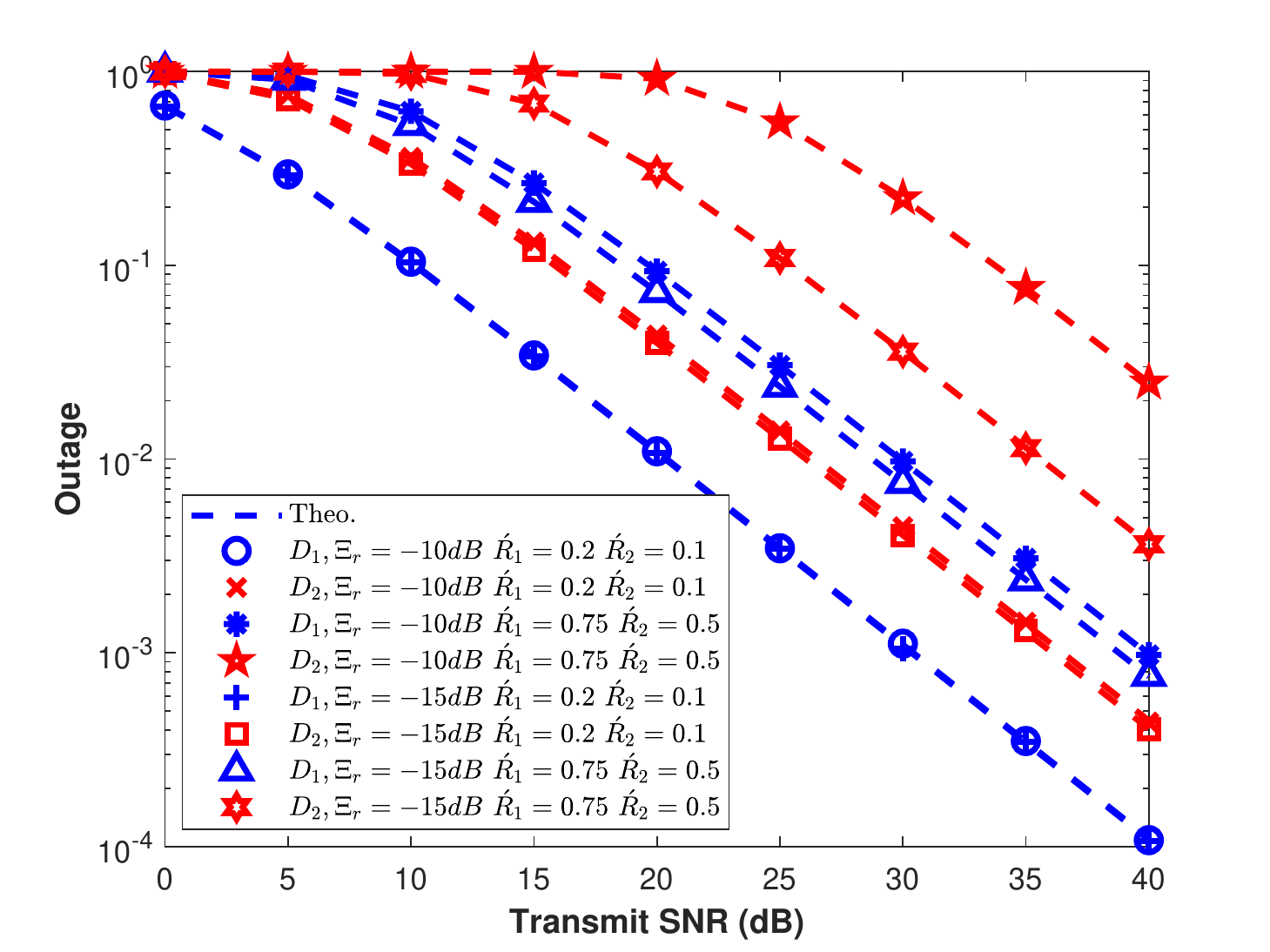}
  \caption{OP of R-DFNOMA vs $\rho_s$ when $\alpha_1=0.9$, $\beta_1=0.2$,  $d_{s,r}=5$, $d_{r,1}=1$ and $d_{r,2}=3$}
\end{figure}
In Fig. 2, EC of users and the ergodic sum-rate of the R-DFNOMA ($C_{sum}=C_1+C_2$) are given for various imperfect SIC effects. Power allocations are assumed to be $\alpha_1=0.9$, $\beta_1=0.2$. As it is expected, imperfect SIC limits the performance of the systems and when it gets higher (i.e., $\Xi_r=\Xi_1=-10dB$), EC of R-DFNOMA becomes worse. The power allocation at the source and relay are chosen as different values for better illustration, otherwise both users' upper bound would be the same.
In Fig. 3, outage performances of the users are presented for the same power allocation coefficients. Target rates of the users are chosen as $\acute{R}_1=0.2$, $\acute{R}_2=0.1$ and $\acute{R}_1=0.75$, $\acute{R}_2=0.5$ as two different QoS requirements. In all NOMA involved systems, the outage performances of the users get better with low QoS requirements (e.g., lower target rates). Likewise, in EC performance, imperfect SIC has a dominant effect on outage performances of users and it may cause users to be in outage with higher QoS.

\begin{figure}[!ht]
  \centering
  \includegraphics[width=8cm]{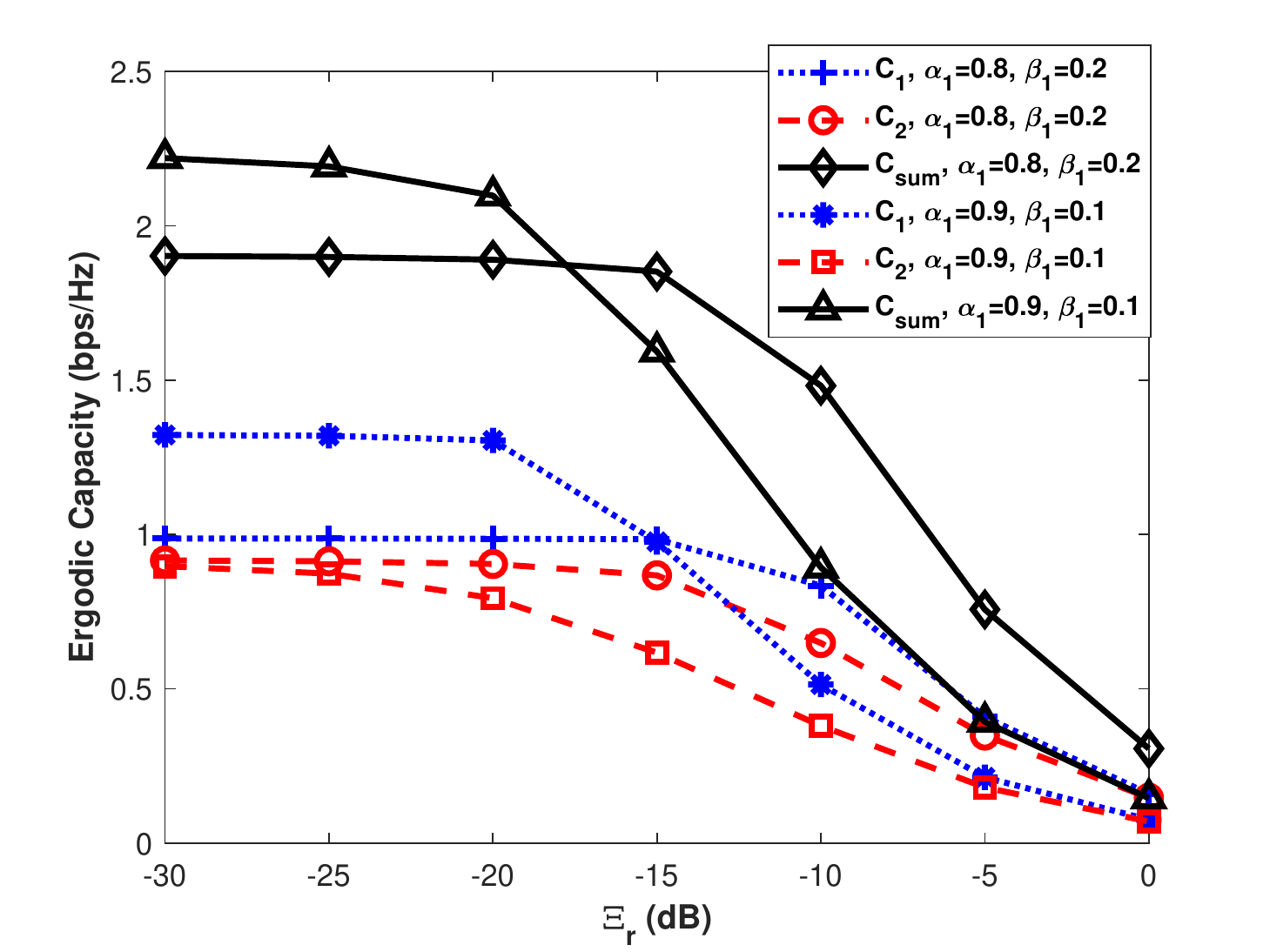}
  \caption{The effect of imperfect SIC on EC of R-DFNOMA when $d_{s,r}=5$, $d_{r,1}=1$, $d_{r,2}=3$ and $\rho_s=20dB$ }
\end{figure}
\begin{figure}[!ht]
  \centering
  \includegraphics[width=8cm]{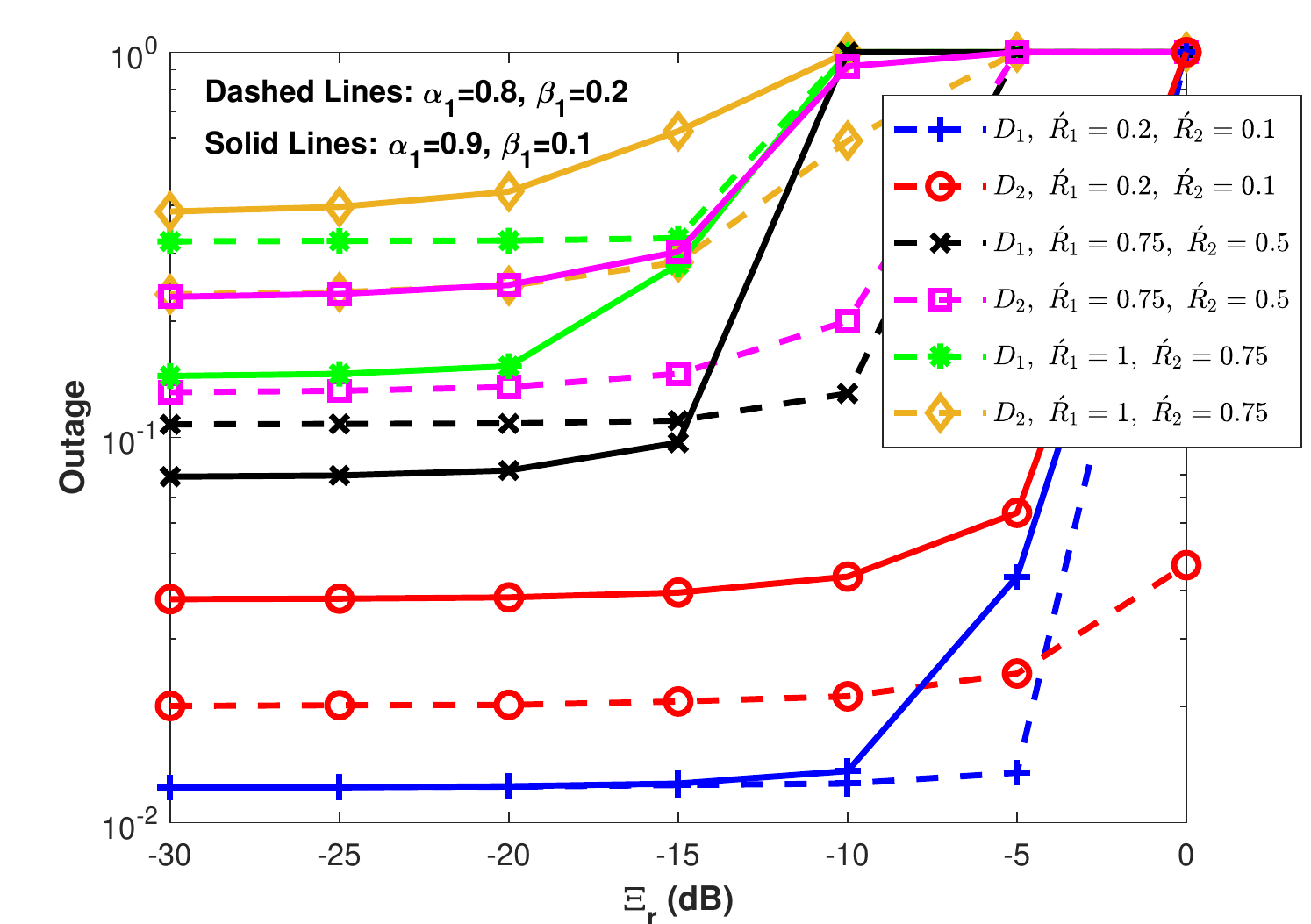}
  \caption{The effect of imperfect SIC on outage of R-DFNOMA when $d_{s,r}=5$, $d_{r,1}=1$, $d_{r,2}=3$ and $\rho_s=20dB$ }
\end{figure}
In order to further investigate the effect of imperfect SIC, we present capacity and outage performance of users with the change of imperfect SIC effect coefficient in Fig. 4 and Fig. 5, respectively. The power allocation coefficients at the source and the relay are assumed to be $\alpha_1=0.8$, $\beta_1=0.2$ and $\alpha_1=0.9$, $\beta_1=0.1$. The result are presented for $\rho_s=20dB$. In bots figures, as expected, with the increase of imperfect SIC effect, users have worse performance in both EC and OP. In Fig. 5, the system will have almost $0.5$ capacity if the imperfect SIC effect is higher than $-5dB$ that is too poor performance for $20dB$ SNR. In Fig. 6, for the same imperfect channel conditions, users are always in outage for all target rates. If more strict QoS requirements are needed (higher target rates), users will be in outage even for lower imperfect SIC effects (e.g., between $-10dB$ and $-5dB$). Nevertheless, all these are expected results for imperfect SIC case and represent more practical/reasonable scenarios than perfect SIC.

\begin{figure}[!ht]
  \centering
  \includegraphics[width=8cm]{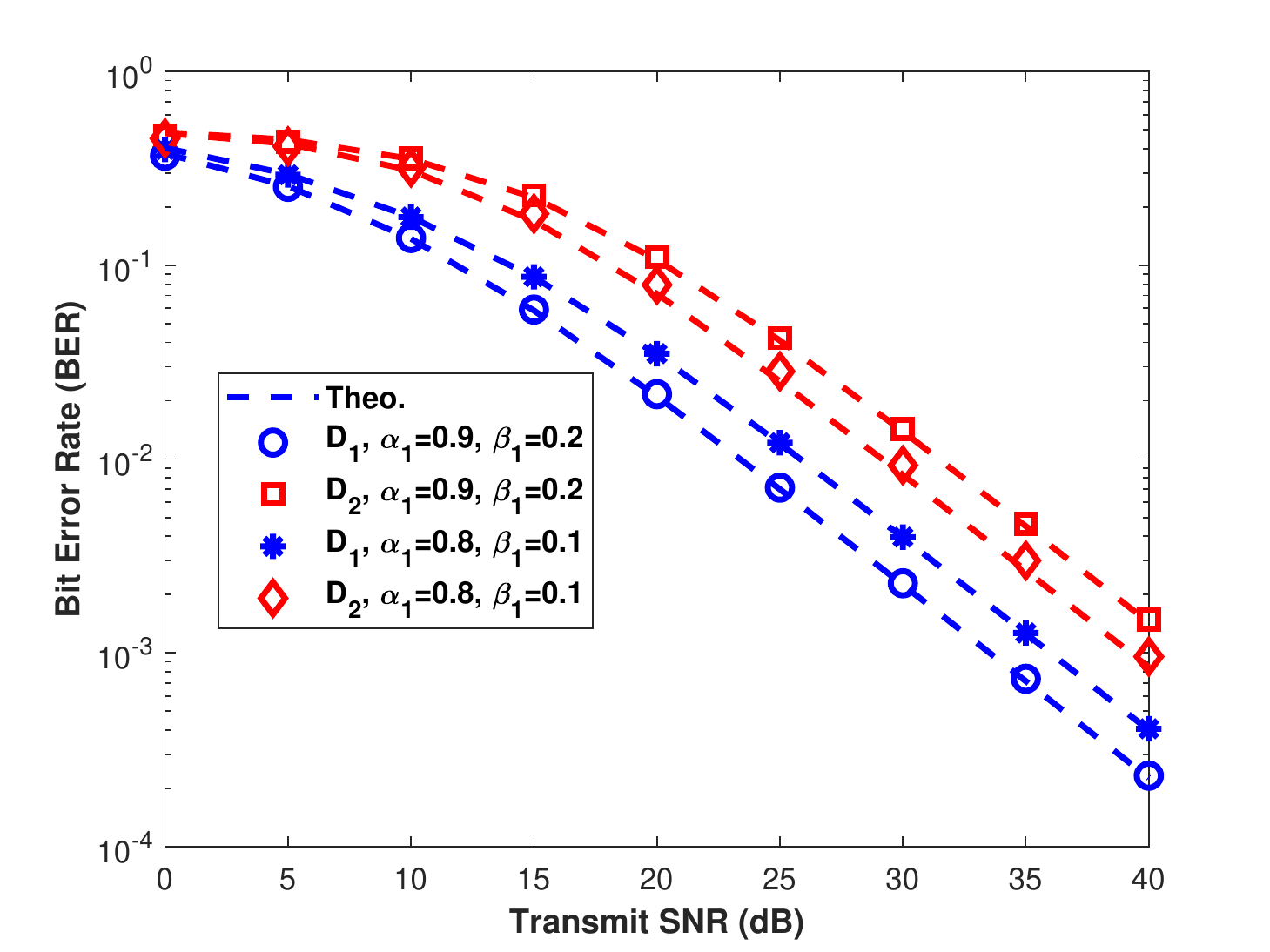}
  \caption{BER of R-DFNOMA vs $\rho_s$ when $d_{s,r}=5$, $d_{r,1}=1$ and $d_{r,2}=3$}
\end{figure}
In Fig. 6, we present error performances of users in R-DFNOMA for $M_1=M_2=4$. In error performance simulations, since an actual modulation/demodulation (i.e., QPSK for $M_i=4$) is implemented contrary to EC and OP simulations, imperfect SIC effect coefficient has not been defined. The erroneous detected symbols during SIC process will show this effect. Thus, we present simulations for two different power allocation pairs (i.e., $\alpha_1=0.8$, $\beta_1=0.2$ and $\alpha_1=0.9$, $\beta_1=0.1$). Just as in previous validations, it is clearly seen in Fig. 6 that derived expressions are perfectly-matched with simulations.

\begin{figure}[!ht]
  \centering
  \includegraphics[width=8cm]{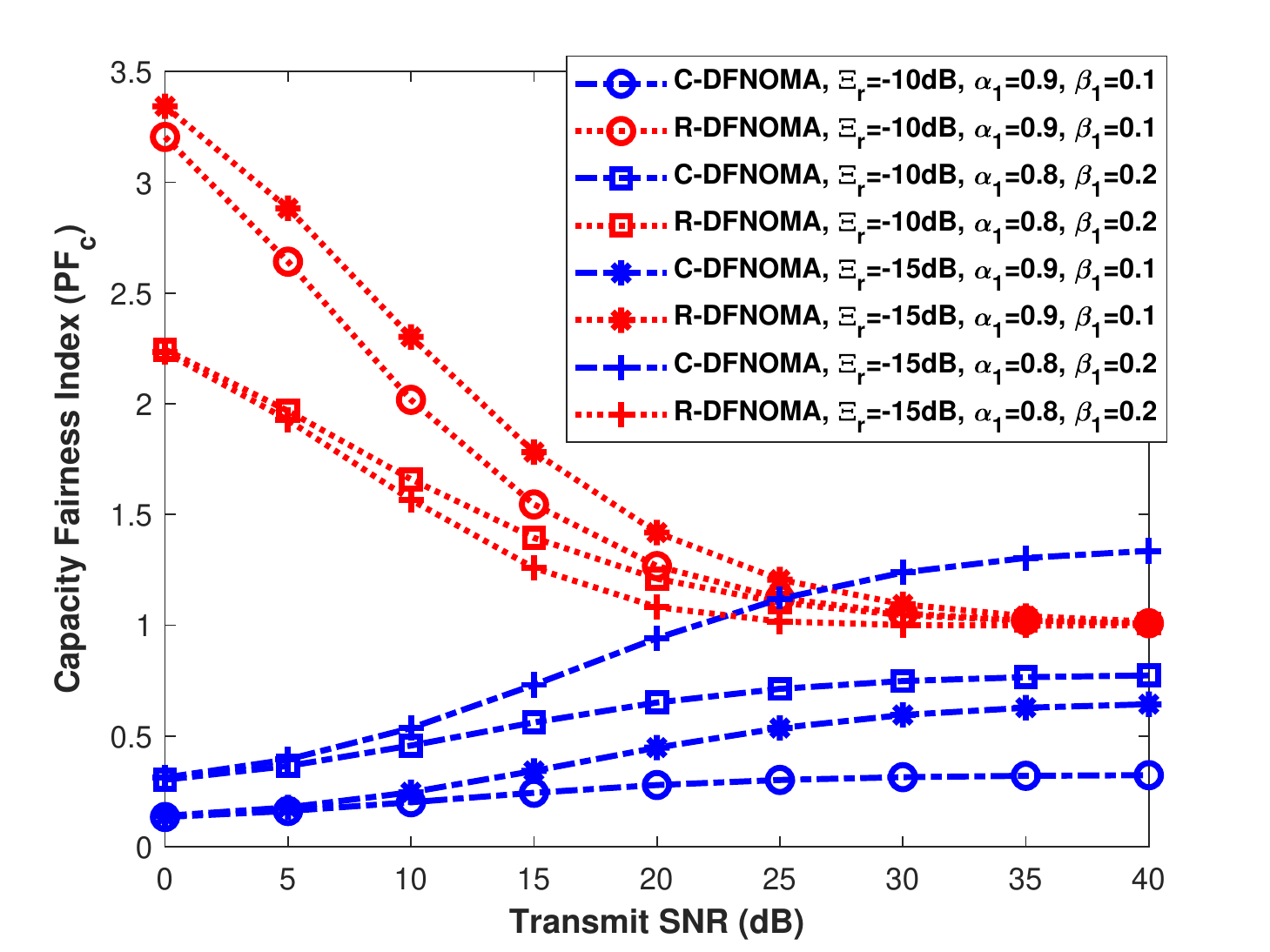}
  \caption{Capacity Fairness Index ($PF_c$) vs $\rho_s$ when $d_{s,r}=5$ and $d_{r,1}=d_{r,2}=2$}
\end{figure}
\begin{figure}[!ht]
  \centering
  \includegraphics[width=8cm]{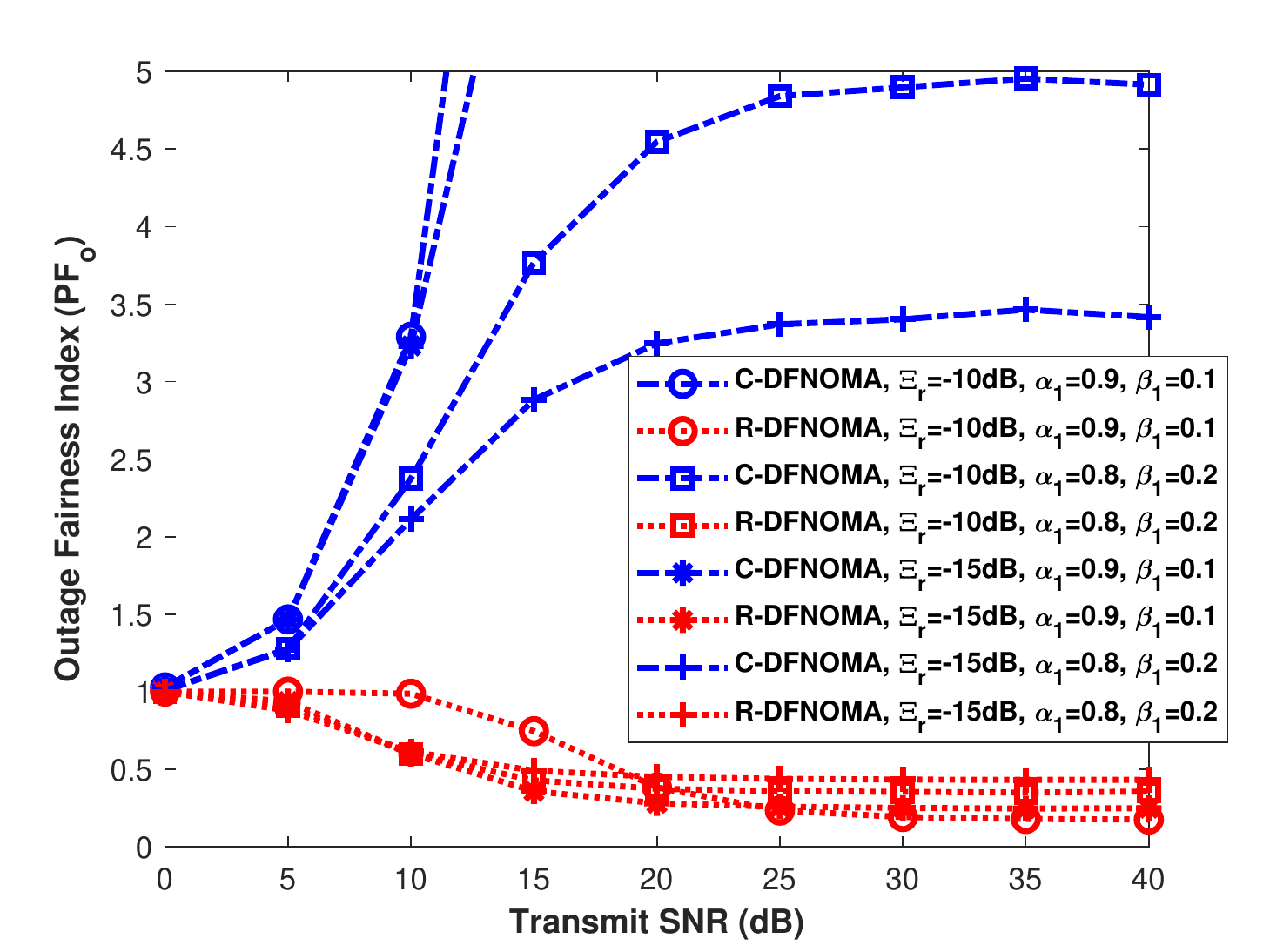}
  \caption{Outage Fairness Index ($PF_o$) vs $\rho_s$ when $d_{s,r}=5$, $d_{r,1}=d_{r,2}=2$ and $\acute{R_1}=\acute{R}_2=0.5$}
\end{figure}
\begin{figure}[!ht]
  \centering
  \includegraphics[width=8cm]{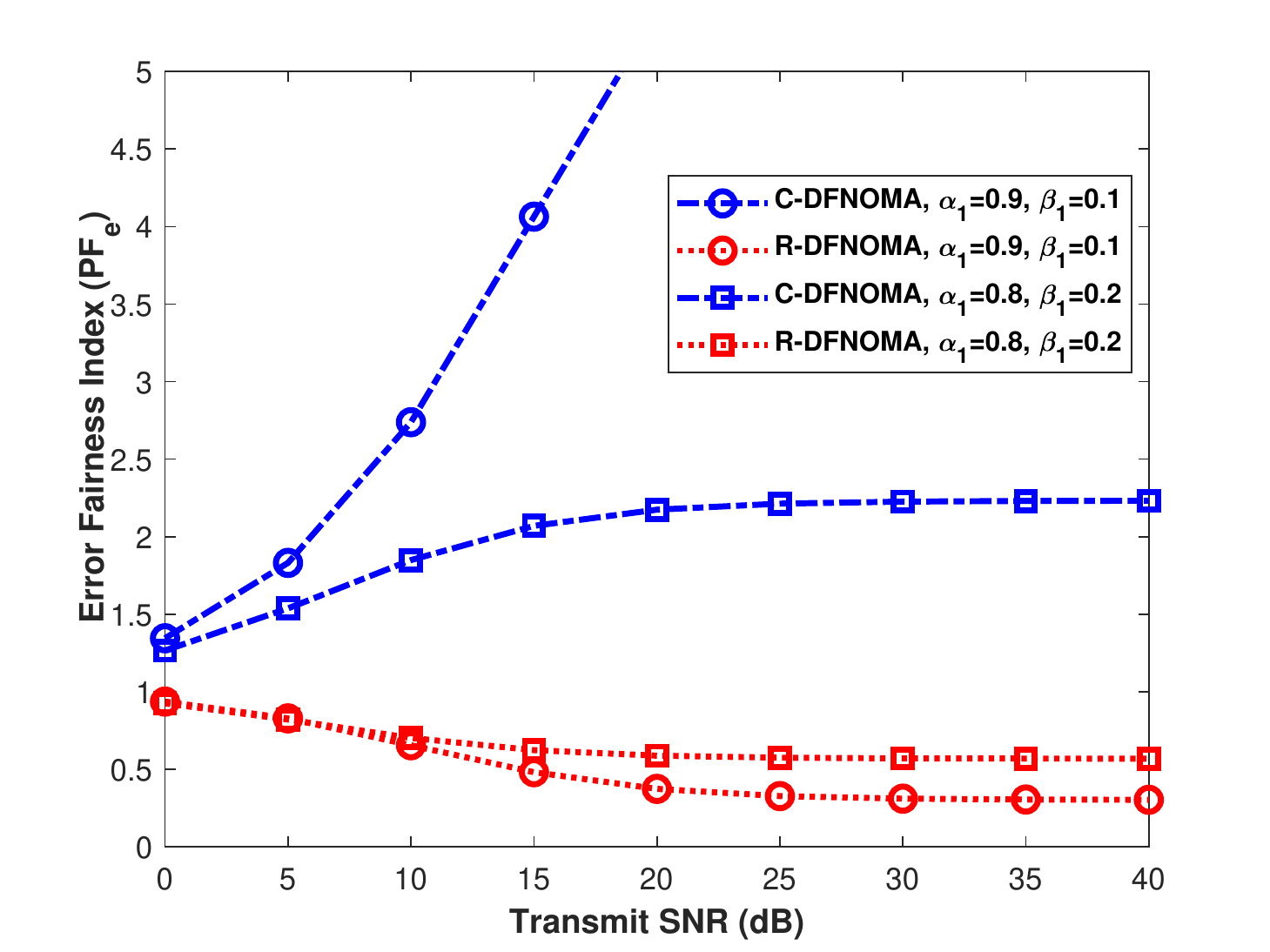}
  \caption{Error Fairness Index ($PF_e$) vs $\rho_s$ when $d_{s,r}=5$ and $d_{r,1}=d_{r,2}=2$}
\end{figure}

\begin{figure*}[!ht]
  \centering
  \includegraphics[width=17cm]{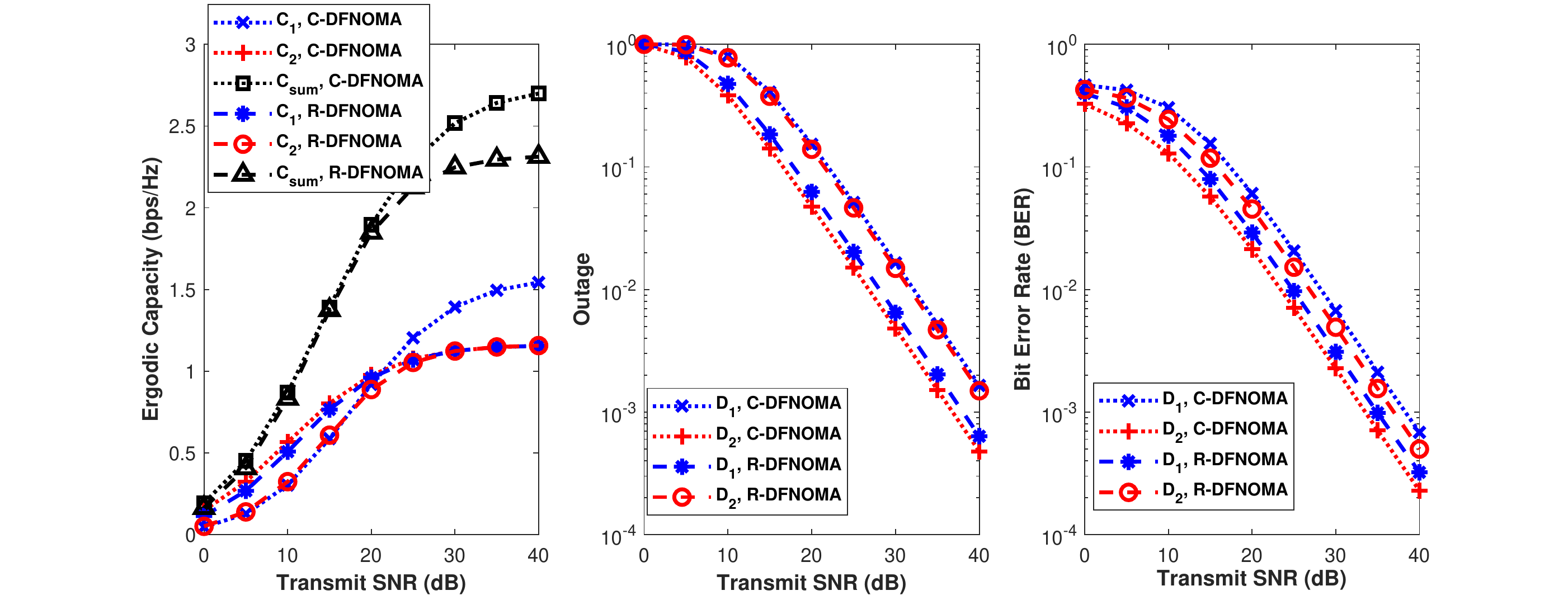}
  \caption{Performance Comparisons between C-DFNOMA and R-DFNOMA vs $\rho_s$ when $d_{s,r}=5$, $d_{r,1}=d_{r,2}=2$, $\alpha_1=0.8$ and $\beta_1=0.2$ a) Capacity when $\Xi_r=-15dB$ b) Outage when $\Xi_r=-15dB$ and $\acute{R}_1=\acute{R}_2=0.5$ c) Bit Error Rate }
\end{figure*}
\subsection{User Fairness}
Contrary to commonly belief, NOMA users do not need to have different channel conditions (stronger and weaker). NOMA users can be chosen among users with similar channel conditions. In this case, user fairness turns out to be more important, since none of them should be served with lower KPI \cite{7906532,8823873}. Thus, user fairness comparisons are more meaningful when the users experience similar channel conditions. To this end, we provide PF index comparisons between R-DFNOMA and C-DFNOMA in Fig. 7- Fig. 9 when $d_{s,r}=5$ and $d_{r,1}=d_{r,2}=2$. In comparisons, two different power allocations scenarios are set $\alpha_1=0.9$, $\beta_1=0.1$ and $\alpha_1=0.8$, $\beta_1=0.2$, respectively. In Fig. 7 and Fig. 8, imperfect SIC effect coefficients are set $\Xi_r=-10dB$ and $\Xi_r=-15dB$. Since the users have similar channel conditions, their target rates are chosen as $\acute{R_1}=\acute{R}_2=0.5$ for $PF_o$. In Fig. 9, modulation orders are set $M_1=M_2=4$. One can easily see that proposed R-DFNOMA provides better user fairness than C-DFNOMA in terms of all KPIs. In proposed R-DFNOMA, users' performance orders are also reversed. In terms of EC in Fig. 7, $D_1$ has higher EC than $D_2$ in R-DFNOMA, hence the fairness index is higher than $1$ whereas it is vice-versa in C-DFNOMA. Nevertheless, R-DFNOMA outperforms C-DFNOMA considering user fairness. For instance, when $\alpha_1=0.9$, $\beta_1=0.1$ and $\Xi_r=-10dB$, $PF_c=1.545$ at $15dB$ SNR in R-DFNOMA which means $D_1$ has $1.545$ times better EC than $D_2$ at $15dB$. However, for the same conditions, $PF_c=0.2433$ in C-DFNOMA, which means $D_2$ has $4.11$ times better EC than $D_1$.\footnote{As discussed in the previous sections, we hereby again note that PF index of any performance metrics have the same meaning for $\kappa$ and $\sfrac{1}{\kappa}$. It only defines which user has better performance.} In addition, it is obvious that R-DFNOMA provides optimum user fairness in high SNR regimes ($PF_c\to1$) whereas it cannot be achieved in C-DFNOMA. Once user fairness is considered in terms of outage performance, the improvement by proposed R-DFNOMA becomes outstanding in Fig. 8. In all considered scenarios, R-DFNOMA achieves about $PF_o=0.5$ for all SNR region which means $D_1$ has $2$ times better OP than $D_2$. However, in C-DFNOMA, with the increase of transmit SNR, user unfairness gets worse. Indeed, in some scenarios user fairness becomes atrocious with the increase of SNR. For instance,  when $\alpha_1=0.9$, $\beta_1=0.1$ and $\Xi_r=-10dB$, $PF_o \to 80$ at $SNR \to 40dB$ which means $D_2$ has $80$ times better OP than $D_1$ although they have same channel conditions. Likewise user fairness comparisons for EC and OP, proposed R-DFNOMA is superior to C-DFNOMA also in terms of error performance in Fig. 9. Furthermore, this improvement is significant in some scenarios. For instance,  when $\alpha_1=0.9$, $\beta_1=0.1$, with the increase of SNR ($\rho_s\to40dB$), $PF_e \to 2$ in R-DFNOMA whereas $PF_e \to 7$ in C-DFNOMA which is a superb gain for data reliability of $D_2$.

If the comparisons between C-DFNOMA and proposed R-DFNOMA are studied only in terms of fairness indexes (e.g., Fig. 7- Fig. 9), the questions may be raised that: Are the user fairness indexes be improved by degrading both of users' performances? or does the R-DFNOMA has worse performance? In order to resolve this concern and to prove that R-DFNOMA achieves better user fairness indexes without degrading users' individual and overall performances, we present performance comparison between C-DFNOMA and R-DFNOMA in Fig. 10. The channel conditions are assumed to be same with above comparisons and the power allocations are fixed as $\alpha_1=0.8$, $\beta_1=0.2$. In capacity and outage comparisons in Fig. 10.a and Fig. 10.b, The imperfect SIC effect coefficient is $\Xi_r=-15dB$ and the target rates of users are  $\acute{R}_1=\acute{R}_2=0.5$. As can be easily from Fig. 10 that in R-DFNOMA, users performances for all KPIs (i.e., EC, OP and BEP) get closer so that user fairness indexes are improved. Besides, when this improvement is achieved, proposed R-DFNOMA does not cause performance decay in either users' or in overall system performances. When the users have similar channel conditions, the overall system performance is limited by the minimum performance achieved within users. To this end,in order evaluate both system (e.g., C-DFNOMA and R-DFNOMA), we can define performance metrics as
\begin{equation}
\begin{split}
    &C^{(m)}=\min\left\{C_1,C_2\right\}, \ m:C-DFNOMA\ or \ R-DFNOMA\\
    &P^{(m)}(out)=\max\left\{P_1(out),P_2(out)\right\}, \\
    &P^{(m)}(e)=\max\left\{P_1(e),P_2(e)\right\}. \\
\end{split}
\end{equation}
and the performance comparisons between R-DFNOMA and C-DFNOMA are given by
\begin{equation}
\begin{split}
    &\max\left\{C^{(C-DFNOMA)},C^{(R-DFNOMA)}\right\}, \\
    &\min\left\{P^{(C-DFNOMA)}(out),P^{(R-DFNOMA)}(out)\right\}, \\
    &\min\left\{P^{(C-DFNOMA)}(e),P^{(R-DFNOMA)}(e)\right\}. \\
\end{split}
\end{equation}
Considering performance comparisons in (46)-(47), we can see from Fig. 10 that, C-DFNOMA and R-DFNOMA has the same performance in terms of theoretical Shannon rate (e.g., Fig. 10.a) where minimum of users' rates are the same in C-DFNOMA and R-DFNOMA. Although it seems that $D_1$ in C-DFNOMA may have higher achievable rate in high SNR region, this should be jointly evaluated with performance of SIC receivers. It is proved in \cite{Lee2019} that when the higher modulation orders (mean higher achievable rate) are implemented, none of the users' symbols cannot be detected and all users have $0.5$ BER performance. On the other hand, R-DFNOMA outperforms C-DFNOMA in terms of outage and BER performances. In Fig. 10.b, $D_2$ has the maximum outage probability in R-DFNOMA (performance limits as given in (46)) whereas $D_1$ has the maximum in C-DFNOMA and $D_2$ in R-DFNOMA has lower OP (better performance as given in (47)) than $D_1$ in C-DFNOMA. Once the same evaluations have been discussed for BER performances, we can easily see that $D_2$ in R-DFNOMA (maximum in R-DFNOMA) has lower BER (better performance as given in (47)) than $D_1$ in C-DFNOMA (maximum in C-DFNOMA).

\begin{figure}[!ht]
  \centering
  \includegraphics[width=8.5cm]{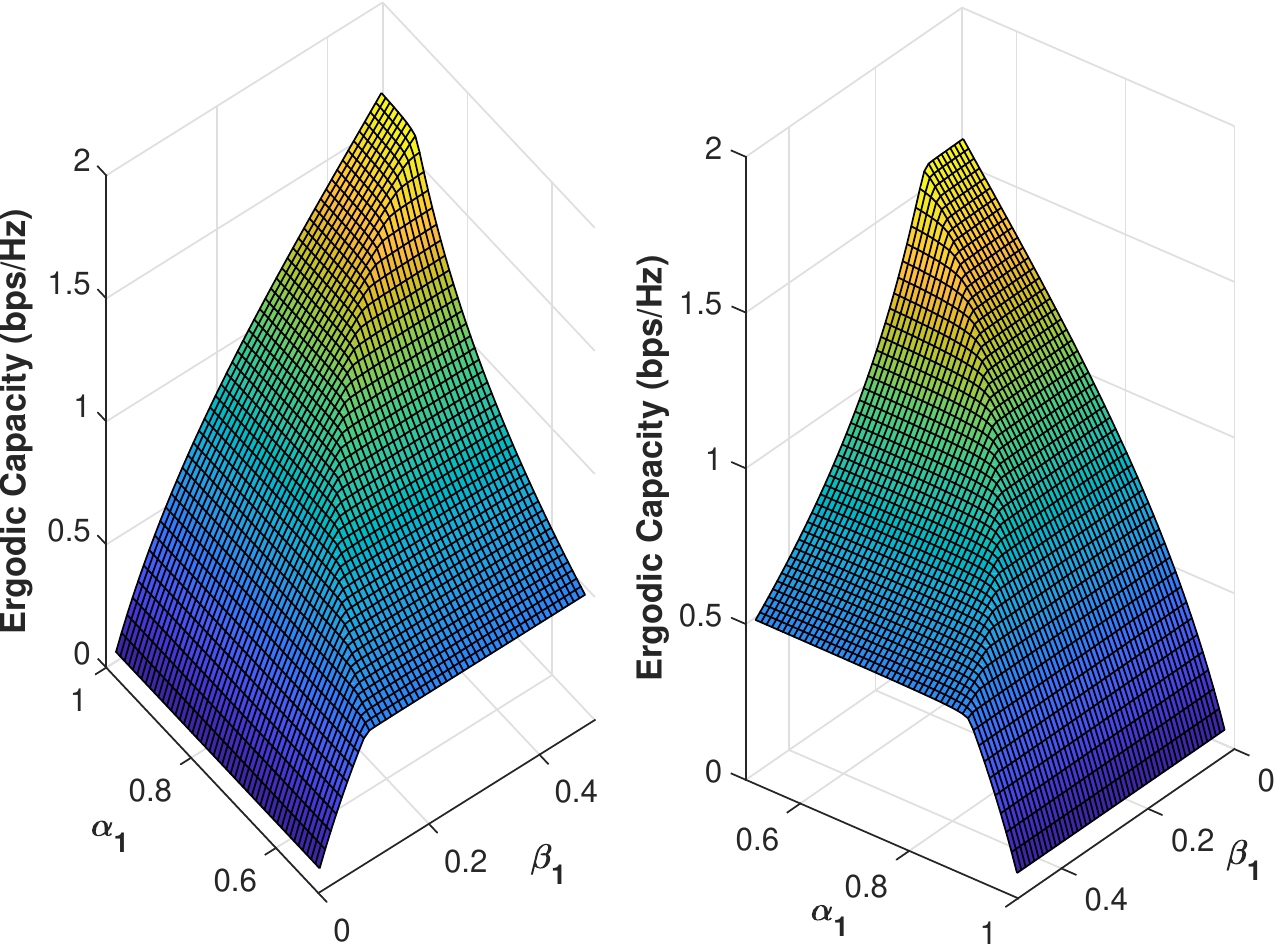}
  \caption{EC performance in R-DFNOMA  vs $\alpha_1$ and $\beta_1$ when $d_{s,r}=5$, $d_{r,1}=d_{r,2}=2$, $\rho_s=30dB$ and $\Xi_r=-10dB$ a) $D_1$ b) $D_2$}
\end{figure}
\begin{figure}[!ht]
  \centering
  \includegraphics[width=8.5cm]{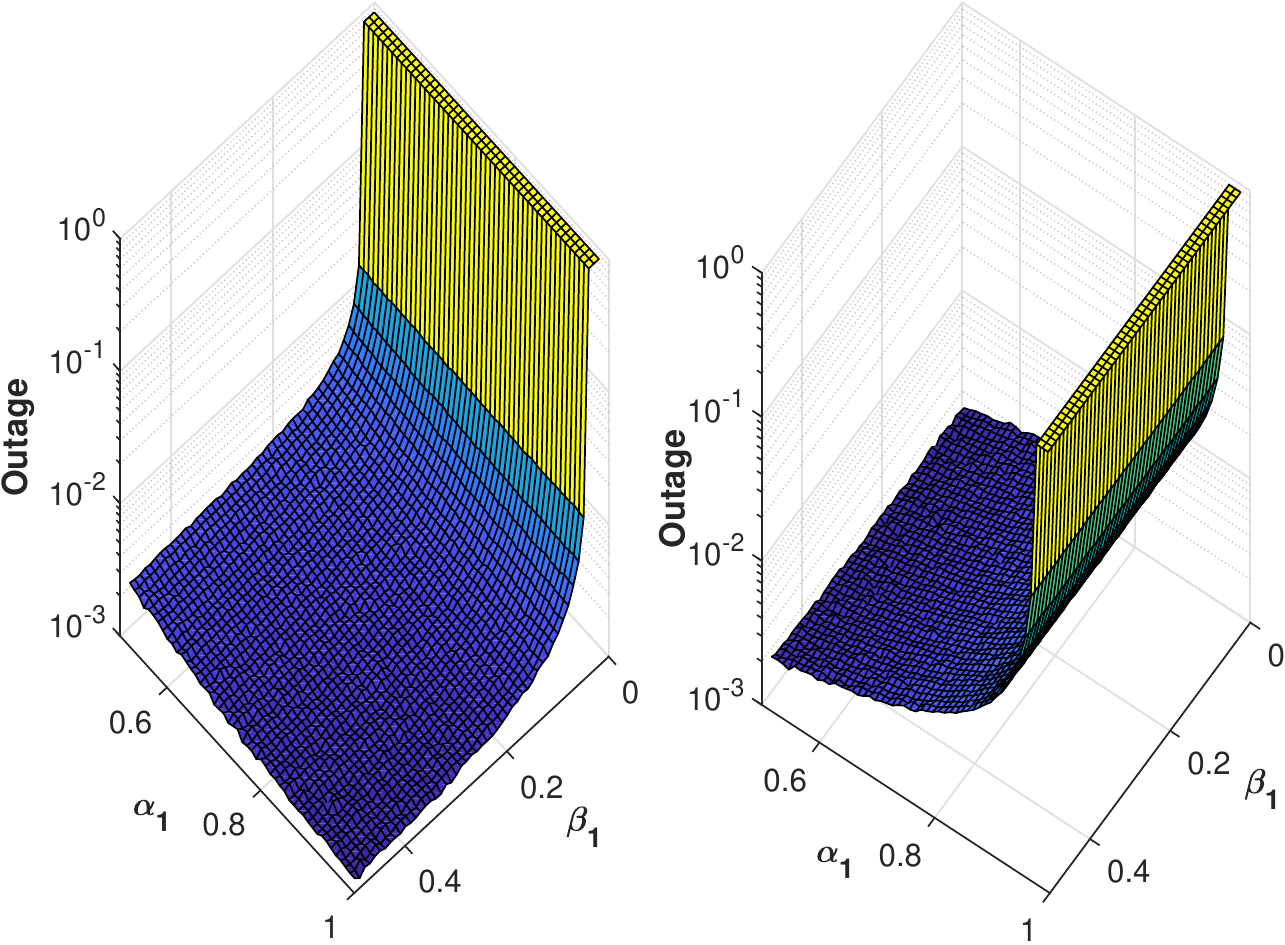}
  \caption{Outage performance in R-DFNOMA  vs $\alpha_1$ and $\beta_1$ when $d_{s,r}=5$, $d_{r,1}=d_{r,2}=2$, $\rho_s=30dB$, $\Xi_r=-10dB$ and $\acute{R}_1=\acute{R}_2=0.2$ a) $D_1$ b) $D_2$}
\end{figure}
\begin{figure}[!ht]
  \centering
  \includegraphics[width=8.5cm]{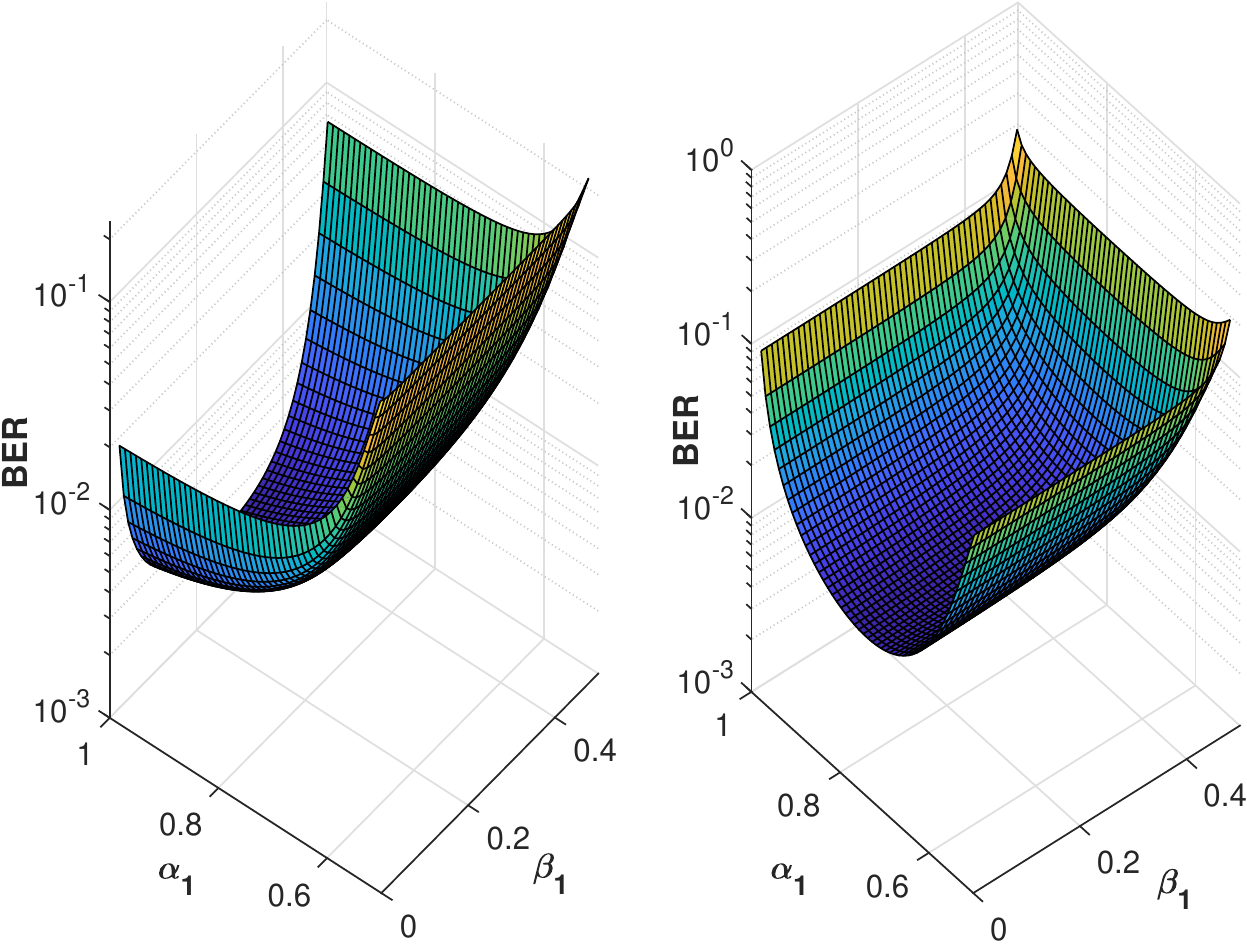}
  \caption{BER performance in R-DFNOMA  vs $\alpha_1$ and $\beta_1$ when $d_{s,r}=5$, $d_{r,1}=d_{r,2}=2$, $\rho_s=30dB$ a) $D_1$ b) $D_2$}
\end{figure}

\subsection{The Effect of Power Allocations}
From above simulations and discussions, it can bee seen that power allocations at the both nodes (source and relay) have a remarkable effect on the performance of R-DFNOMA so that on the user fairness. Thus, in order to reveal this effect, we present EC, OP and BER performances of users with the respect to power allocations ($\alpha_1$ and $\beta_1$) in Fig. 11 - Fig. 13. The channel conditions are assumed to be $d_{s,r}=5$ and $d_{r,1}=d_{r,2}=2$. The imperfect SIC effect is $\Xi_r=-10dB$ assumed in EC and OP comparisons and transmit SNR $\rho_s=30dB$ is assumed in all figures. Target rates of users are assumed to be equal to $\acute{R}_1=\acute{R}_2=0.2$. $M_1=M_2=4$ is set in BER simulations. As expected, both power allocation coefficients (source and relay) have reverse effects on the performances of users. Increasing/decreasing $\alpha_1$ and/or $\beta_1$ provides performance gain for one of the user and causes a decay for the other vice-versa. In terms of EC in Fig. 11, increase in $\alpha_1$ and/or $\beta_1$ provide better EC for $D_1$ and lower EC for $D_2$. The same discussions are also valid for outage performances of users in Fig. 11. Nevertheless, it is noteworthy that increasing/decreasing any power allocation coefficient too much causes one of the users to be in always outage. The same performance trends can be seen for also BER in Fig. 13. However, since an actual modulation/demodulation is implemented, the effect of power allocation on the SIC performance is more clear in Fig. 13. For instance, increasing $\beta_1$ means that higher power is allocated to $D_1$ symbols in the second phase thereby better error performance is expected. Nevertheless, one can easily see that increasing $\beta_1$ too much causes a decay in error performance of $D_1$ along with error performance of $D_2$. This is explained as follows. Since the $D_1$ has to implement SIC in the second phase, increasing $\beta_1$ causes erroneous SIC more likely and this pulls down the error performance of $D_1$. Based on above discussions, it is clear that power allocation affects users' performances reversely (gain for one and decay for the other), thus it is not possible to define a optimum power allocation which offers the best performances for both users. Nevertheless, by considering user fairness, a sub-optimum power allocation can be obtained.

To this end, user fairness comparisons between R-DFNOMA and C-DFNOMA with respect to power allocations ($\alpha_1$ and $\beta_1$) in Fig. 14 - Fig. 16 for the same conditions above comparisons. Based on user fairness for EC in Fig. 14, although $PF_c$ changes within larger range (max. $5$) in R-DFNOMA, R-DFNOMA provides better user fairness (close to $1$), whereas user fairness in C-DFNOMA is close to $1.5$, for most of power the allocation pairs. In Fig. 15, user fairness is presented in terms of OP when the users have same QoS requirement ($\acute{R}_1=\acute{R}_2=0.2$). It can bee seen that users in R-DFNOMA have similar outage performance even if power allocation coefficients change. In R-DFNOMA, one of the users may have maximum $6$ times better outage performance than the other. However, this unfairness may raise to $50$ times better performance in C-DFNOMA. Lastly, user fairness in terms of error performance is presented in Fig. 16. Similar discussions can be seen for also user fairness in terms of BER from Fig. 16. The fairness index may have the maximum $3.5$ in R-DFNOMA whereas the value of $80$ in C-DFNOMA. This shows that users in R-DFNOMA have similar data reliability, however in C-DFNOMA, one of the users may outperform $80$ times the other. Based on the results between Fig. 14 and Fig. 16, R-DFNOMA is much more robust to change in power allocation coefficients in terms of user fairness. For the given channel conditions, by considering the user fairness in terms of all KPIs, the optimum power allocation pairs in R-DFNOMA could be defined $\alpha_1\approx0.85$ and $\beta_1\approx0.15$ where users have same performance so that user fairness becomes very close to $1$.
\begin{figure}[!ht]
  \centering
  \includegraphics[width=8.5cm]{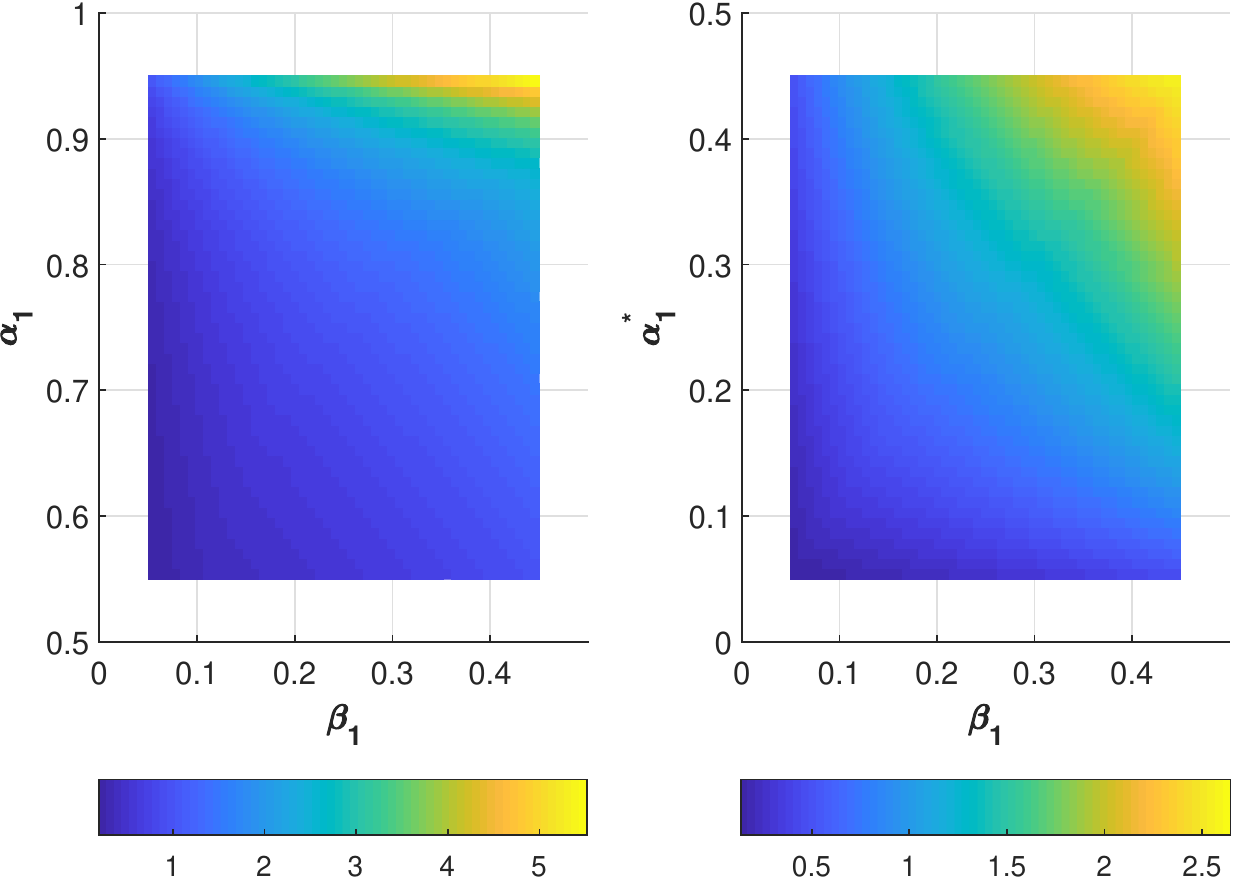}
  \caption{Capacity Fairness Index ($PF_c$) vs $\alpha_1$ and $\beta_1$ when $d_{s,r}=5$, $d_{r,1}=d_{r,2}=2$, $\rho_s=30dB$ and $\Xi_r=-10dB$ a) R-DFNOMA b) C-DFNOMA}
\end{figure}
\begin{figure}[!ht]
  \centering
  \includegraphics[width=8.5cm]{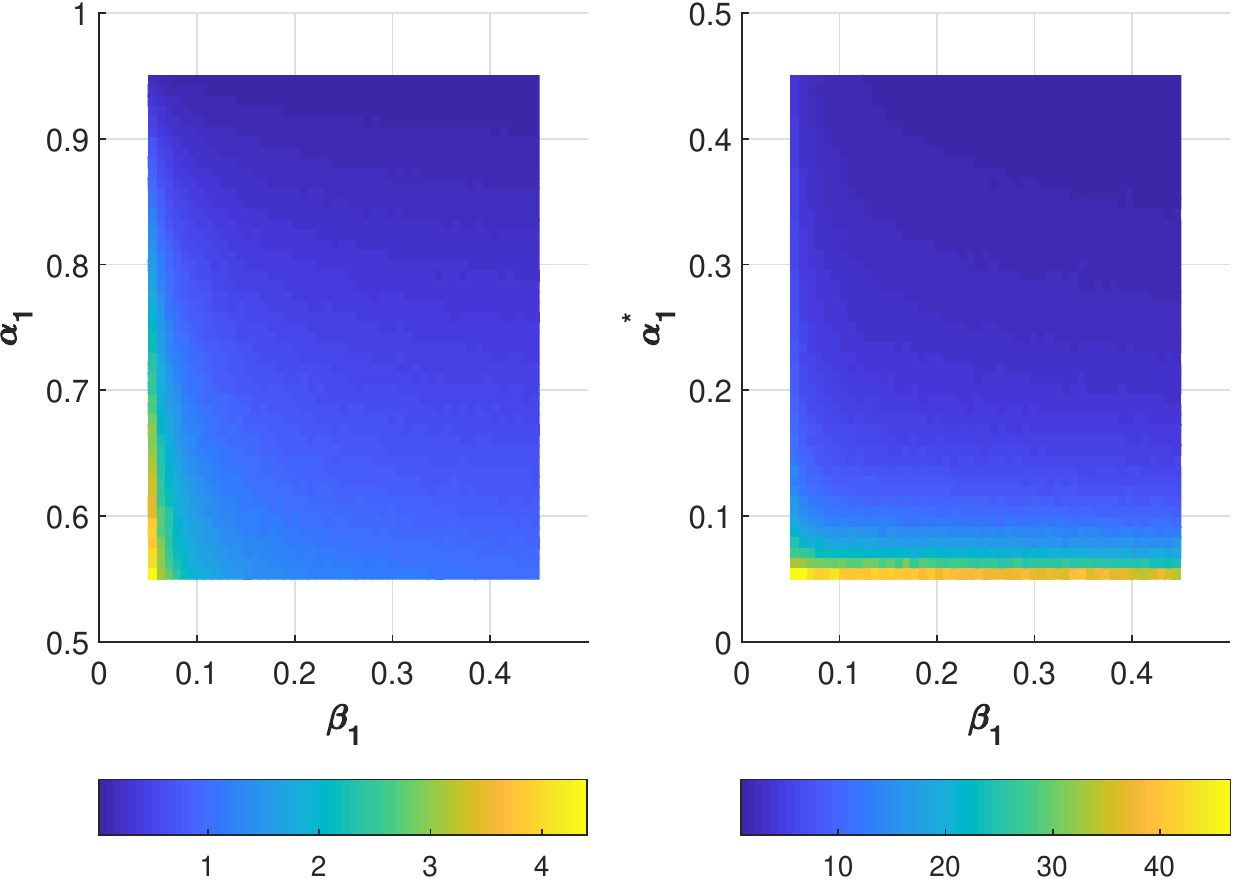}
  \caption{Outage Fairness Index ($PF_o$) vs $\alpha_1$ and $\beta_1$ when $d_{s,r}=5$, $d_{r,1}=d_{r,2}=2$, $\rho_s=30dB$, $\Xi_r=-10dB$ and $\acute{R}_1=\acute{R}_2=0.2$ a) R-DFNOMA b) C-DFNOMA}
\end{figure}
\begin{figure}[!ht]
  \centering
  \includegraphics[width=8.5cm]{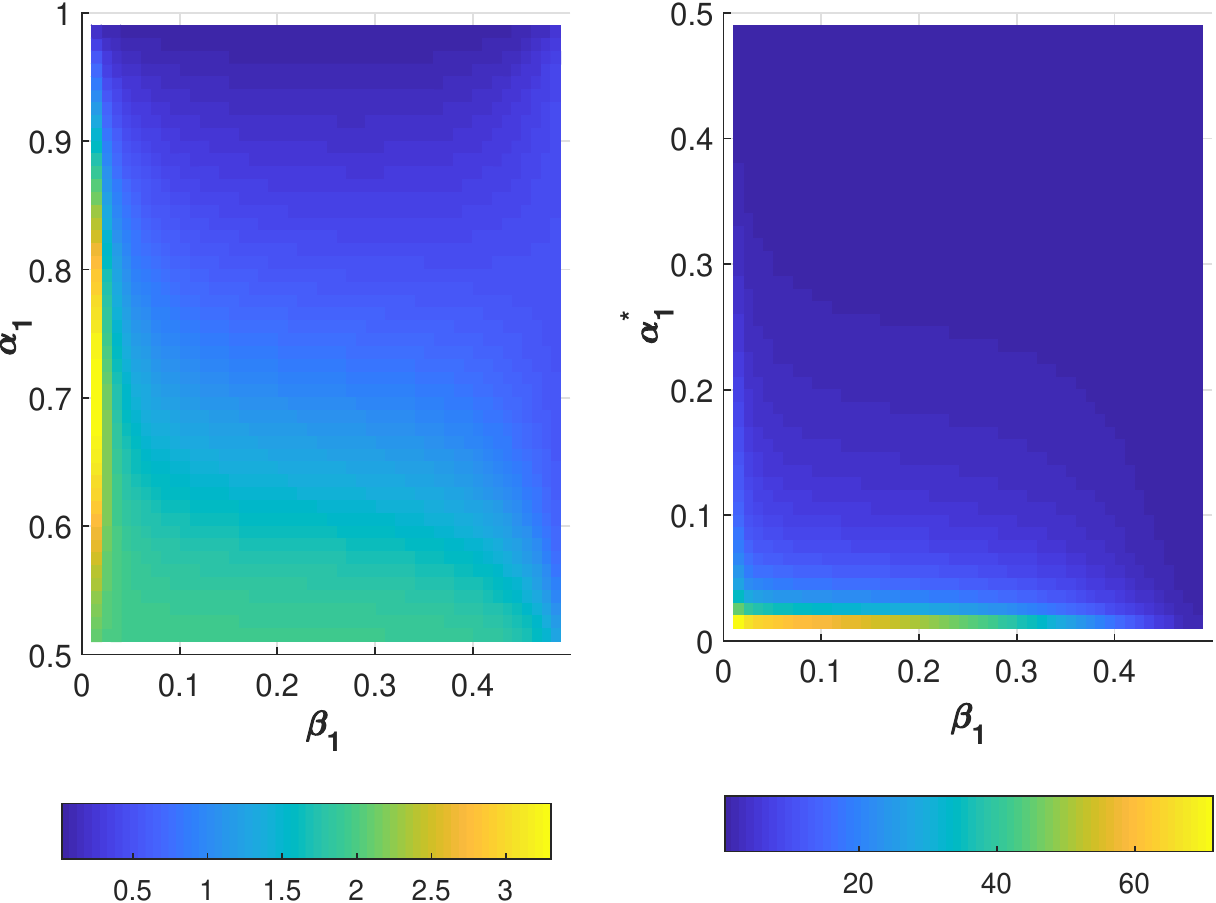}
  \caption{Error Fairness Index ($PF_e$) vs $\alpha_1$ and $\beta_1$ when $d_{s,r}=5$, $d_{r,1}=d_{r,2}=2$ and $\rho_s=30dB$ a) R-DFNOMA b) C-DFNOMA}
\end{figure}
\section{Conclusion}
In this paper, we introduce reversed decode-forward relaying NOMA (R-DFNOMA) to improve user fairness in conventional DFNOMA (C-DFNOMA). We consider imperfect SIC at detections and according to imperfect SIC effect and we re-define SINRs at the nodes to provide a more practical scenario. With the imperfect SIC, we investigate the performance of proposed R-DFNOMA and derive closed-form expressions for ergodic capacity (EC), outage probability (OP) and bit error probability (BEP). Then, in order to emphasize user fairness, we define fairness indexes for all KPIs (i.e., EC, OP and BEP). With the extensive simulations, derived expressions are validated and it is proved that R-DFNOMA outperforms significantly C-DFNOMA in terms of user fairness when the users have similar channel conditions. Moreover, user fairness is investigated with the change of power allocation coefficients and optimum power allocation is discussed under user fairness constraint. Based on the results and discussions, it is revealed that reversed power allocation at the source and changing SIC order at the relay can provide better user fairness and the system will be more robust to power allocation. This promising result shows that reversed networks can be implemented in other NOMA and cooperative involved systems. To this end, analysis and discussions for DF relaying in this paper can be extended for other strategies. Hence, reversed power allocation in DFNOMA with full-duplex, DFNOMA with energy harvesting and DFNOMA with direct links are seen as future research topics.
\bibliographystyle{IEEEtran}
\bibliography{improved_uf_df_nomav2}

%
%
\begin{IEEEbiography}[{\includegraphics[width=1in,height=1.25in,clip,keepaspectratio]{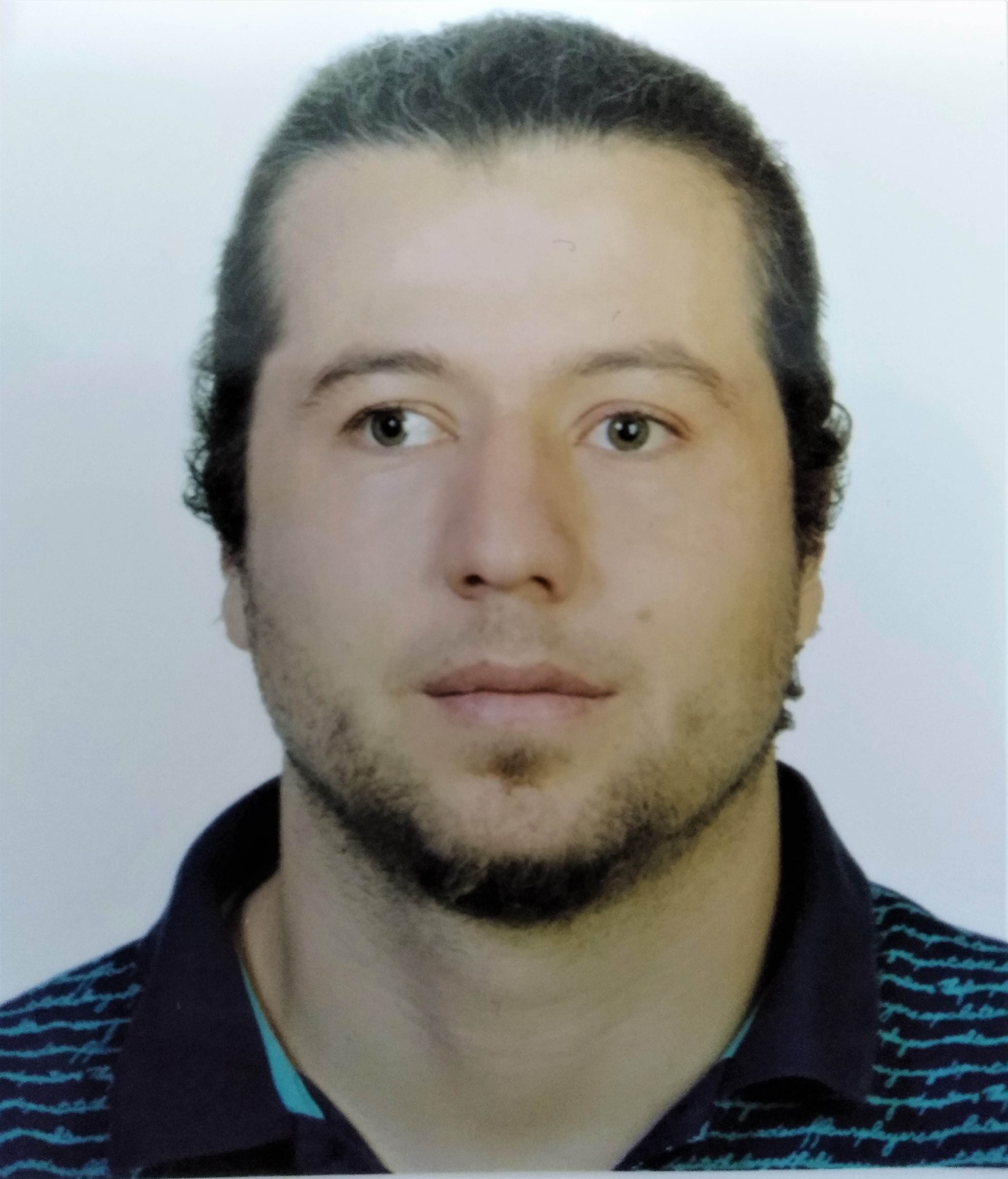}}]{Ferdi Kara}received B.Sc. (with  Hons.) in electronics and communication engineering from Suleyman Demirel University, Turkey in 2011. He received M.Sc. and PhD degrees in electrical and electronics engineering from Zonguldak Bulent Ecevit University, Turkey, in 2015 and 2019, respectively. Since 2011, he works with Wireless Communication Technologies Research Laboratory (WCTLab), where he now holds Senior Researcher position. He serves as regular reviewer for reputed IEEE journals and conferences. He was awarded with Exemplary Reviewer Certificate  by IEEE Communications Letters in 2019. His research area spans within wireless communications specified with NOMA, MIMO systems, cooperative communication, index modulations, energy harvesting, and machine learning in physical communication.
\end{IEEEbiography}
\begin{IEEEbiography}[{\includegraphics[width=1in,height=1.25in,clip,keepaspectratio]{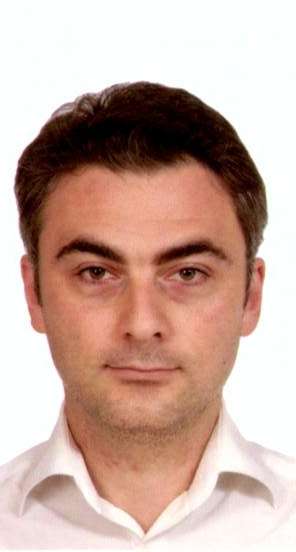}}]{Hakan Kaya}received B.Sc., M.Sc., and Ph.D degrees in all electrical and electronics engineering from Zonguldak Karaelmas University, Turkey, in 2007, 2010, and 2015, respectively. Since 2015, he has been working at Zonguldak Bulent Ecevit University as Assistant Professor and the head of the Wireless Communication Technologies Research Laboratory (WCTLab). His research interests are cooperative communication, NOMA, turbo coding and machine learning.
\end{IEEEbiography}

\end{document}